\documentclass[useAMS, usenatbib]{mn2e}
\usepackage{afterpage}
\usepackage{tensor}
\usepackage{graphicx}
\usepackage{subfigure}
\usepackage{amsmath, amssymb, amsfonts}

\newcommand \rin {r_{\rm in}}
\newcommand \rout {r_{\rm out}}

\newcommand \rhoe {\rho_{\rm e}}
\newcommand \divE {\nabla \cdot \vec{E}}
\newcommand \divB {\nabla \cdot \vec{B}}
\newcommand \curlE {\nabla \times \vec{E}}
\newcommand \curlB {\nabla \times \vec{B}}
\newcommand \E {\vec{E}}
\newcommand \B {\vec{B}}
\newcommand \J {\vec{J}}
\newcommand \rlc {R_{\rm LC}}

\newcommand \dd {\partial}
\newcommand \beq {\begin{equation}}
\newcommand \eeq {\end{equation}}

\newcommand \lp {\left(}
\newcommand \rp {\right)}
\newcommand \lb {\left\{}
\newcommand \rb {\right\}}
\newcommand \ls {\left[}
\newcommand \rs {\right]}

\newcommand{\mtx}[1]{\mathsf{#1}}

\let\oldhat\hat
\renewcommand{\vec}[1]{\boldsymbol{#1}}
\renewcommand{\hat}[1]{\oldhat{\mathbf{#1}}}

\title[\textsc{PHAEDRA}: a new spectral code]{Introducing \textsc{PHAEDRA}: a new spectral code for simulations of relativistic magnetospheres}

\author[Parfrey et al.]{Kyle Parfrey$^{1,2}$\thanks{E-mail: kyle@astro.columbia.edu}, Andrei M. Beloborodov$^{3}$, and Lam Hui$^{2,3}$\\
$^{1}$Astronomy Department, Columbia University, 550 West 120th Street, New York, NY 10027, USA\\
$^{2}$Institute for Strings, Cosmology, and Astroparticle Physics (ISCAP), Columbia University, New York, NY 10027, USA\\
$^{3}$Physics Department and Columbia Astrophysics Laboratory, Columbia University, 538 West 120th Street, New York, NY 10027, USA}
\begin{document}

\maketitle
\label{firstpage}
\begin{abstract}
We describe a new scheme for evolving the equations of force-free electrodynamics, the vanishing-inertia limit of magnetohydrodynamics. This pseudospectral code uses global orthogonal basis function expansions to take accurate spatial derivatives, allowing the use of an unstaggered mesh and the complete force-free current density. The method has low numerical dissipation and diffusion outside of singular current sheets. We present a range of one- and two-dimensional tests, and demonstrate convergence to both smooth and discontinuous analytic solutions. As a first application, we revisit the aligned rotator problem, obtaining a steady solution with resistivity localised in the equatorial current sheet outside the light cylinder.
\end{abstract}


\section{Introduction}

We present a code for simulations of force-free electrodynamics: \textsc{phaedra} (Pseudospectral High-Accuracy ElectroDynamics for Relativistic Astrophysics). The systems we study are `force-free' in the sense that the Lorentz force density vanishes everywhere, because the electromagnetic fields are strong enough that hydrodynamic forces can be neglected, resulting in self-balancing electromagnetic fields. Relativistic force-free electrodynamics has long been recognised as the appropriate limit for describing the magnetospheres of neutron stars and black holes, yet only recently has a concerted effort begun to study it with direct numerical simulation. 

This infinite-magnetisation, or vanishing-inertia, limit is the appropriate one for the magnetospheres of pulsars \citep*{Goldreich:1969p4440, Contopoulos:1999p3926, Gruzinov:2006p4157, Spitkovsky:2006p752}, and magnetars, whose persistent and transient high-energy emission may be due to the distortion, reconnection, and dissipation of force-free fields \citep{ThompDunc:1995, Lyutikov:2006, Beloborodov:2009}. Force-free electrodynamics is the standard tool to study the extraction of rotational energy from black holes \citep{Blandford:1977p4121, MacDonald:1982p4117}, where the magnetic field is thought to be supplied by a conducting accretion disc. The natural self-collimation of electromagnetic fields make them attractive candidates for explaining relativistic jets in quasars and active galactic nuclei, whose high Lorentz factors suggest low baryon loading and electromagnetic dominance \citep{Blandford:1976p4168}; similar Poynting jets may be responsible for gamma-ray bursts \citep{Meszaros:1997p4192}. An argument can be made that all ultra-relativistic outflows are essentially electromagnetic, rather than gas dynamical \citep{Blandford:2002p4143}. 

In any case, it is clear that there exists a wealth of challenging problems for the field of electrodynamic numerical simulation. Direct time-dependent simulation is valuable because it permits the study of general realistic initial-value problems, without the restrictions, like self-similarity or stationarity, that are often necessary in analytical models, and because it naturally tests the stability of field configurations, a question often unanswered by steady-state numerical work.

Several time-dependent force-free electrodynamics codes exist, both those using finite differences \citep{Spitkovsky:2006p752, Kalapotharakos:2009p3903, 2010PhRvD..82d4045P}, and those that take a finite volume, or Godunov, approach \citep*{Komissarov:2004p747, Cho:2005p4091, Asano:2005p4081, McKinney:2006p3434, Yu:2011p4007}. Our numerical scheme is entirely different, and complementary, being based on orthogonal basis function expansions.

Previous codes have large numerical dissipation or diffusion, introduced either because they do not maintain $\E\cdot\B=0$ self-consistently or through the intrinsic diffusivity of the method, while force-free problems often demand long simulations, as the fields may evolve over many wave-crossing times. It is desirable to have a method which can be run for long times without intrinsic dissipation, captures discontinuities, and accurately describes fast dynamics.

The crucial question one asks of a force-free configuration is that of its stability, the onset of instability commonly leading to a dramatic rearrangement of a magnetosphere, sometimes involving explosive reconnection. Spectral calculations tend to  have less numerical noise than those of comparable finite-difference or finite-volume (`local') schemes; this noise can erroneously trigger instability. In a study of Sweet-Parker reconnection, the spectral magnetohydrodynamics (MHD) code is found to be largely immune to the secondary island formation, caused by a tearing-mode instability, that is found using local methods for the same problem \citep{SNg:2011p3295}. 

In this paper, we describe a code for axisymmetric simulations, in flat space-time. It has been designed in such a way as to be extensible  with minimal restructuring to a fully three-dimensional setting, in curved space-time.

Force-free electrodynamics, and its relation to relativistic MHD, is discussed in Section~\ref{sec:ffe}. Section~\ref{sec:nummethod} contains a detailed description of the code and its practical implementation, including some background on each of its components. A range of one- and two-dimensional test problems are presented in Section~\ref{sec:tests}, including convergence tests in realistic scenarios. The aligned rotator is examined in Section~\ref{sec:pulsar}. Finally, we discuss our results in Section~\ref{sec:discuss}, and outline some promising areas of future research.

Note that in Section~\ref{sec:nummethod} we distinguish between the contravariant, $F^i$, and covariant, $F_i$, components of a vector field, while when we discuss results in Sections~\ref{sec:tests} and \ref{sec:pulsar} we refer only to the components in an orthonormal basis, also written $F_i$.

\section{Force-free electrodynamics}
\label{sec:ffe}
The system of force-free electrodynamics is the vanishing-inertia, or, equivalently, ultra-relativistic, limit of magnetohydrodynamics. The latter can be written as 
\beq
\nabla_{\mu} \tensor[^*]{F}{^{\mu\nu}} = 0, 
\label{eq:maxwell1}
\eeq
\beq
\nabla_{\mu} T^{\mu\nu} = \nabla_{\mu} \lp T_{\rm (m)}^{\mu\nu} + T_{\rm (f)}^{\mu\nu} \rp = 0,
\label{eq:energyconserve}
\eeq
where $\tensor[^*]{F}{^{\mu\nu}}$ is the Maxwell tensor (Hodge dual of the Faraday tensor $F^{\mu\nu}$), and $T_{\rm (m)}^{\mu\nu}$ and $T_{\rm (f)}^{\mu\nu}$ are the energy-momentum tensors of the matter and electromagnetic fields, respectively. If the matter contribution to $T^{\mu\nu}$ can be neglected, equation~(\ref{eq:energyconserve}) simplifies to
\beq
\nabla_{\mu} T_{\rm (f)}^{\mu\nu} = 0.
\label{eq:fieldconserve}
\eeq
Combining
\beq
T_{\rm (f)}^{\mu\nu} = \tensor{F}{^\mu_\alpha}\tensor{F}{^{\nu\alpha}} - \frac{1}{4}\lp \tensor{F}{_{\alpha\beta}}\tensor{F}{^{\alpha\beta}} \rp \tensor{g}{^{\mu\nu}},
\eeq
where $g_{\mu\nu}$ is the metric tensor of space-time, with the inhomogeneous Maxwell equations,
\beq
\nabla_\nu F^{\mu\nu} = J^\mu,
\label{eq:maxwell2}
\eeq
one finds that equation~(\ref{eq:fieldconserve}) becomes
\beq
F_{\mu\nu} J^\nu = 0,
\label{eq:covforcefree}
\eeq
which states that the Lorentz force density vanishes \citep[e.g.][]{Komissarov:2002p721}. Equation~(\ref{eq:covforcefree}) can alternatively be derived by postulating, in a frame in which the electric and magnetic fields are parallel, that the electric field vanishes and the current flows along the magnetic field.

Standard MHD codes can become inaccurate, and even crash, when the plasma's magnetisation is large, because the numerical error in the electromagnetic energy density becomes comparable to the matter energy density \citep*{Gammie:2003p722, Komissarov:2004p4161}. They also require the (usually poorly constrained) matter distribution to be set at the beginning of the simulation, and maintained throughout, sometimes by ad-hoc matter injection. Force-free codes do not experience these difficulties.

We move now to a $3+1$ space-time point of view. Equation~(\ref{eq:covforcefree}) becomes
\beq
\rhoe \E + \J \times \B = 0,
\label{eq:forcefree}
\eeq
where $\E$ and $\B$ are the electric and magnetic fields, and $\J$ the current density. We can see that
\beq
\E \cdot \B = 0;
\label{eq:edotb}
\eeq
the electric field is `degenerate.' This condition, together with $\divB = 0$, implies that the system of electromagnetic fields has only four independent components. Likewise,
\beq
\E \cdot \J = 0;
\eeq
there is no Joule heating, the system is dissipationless, and formally (conditionally) hyperbolic.

The evolutionary Maxwell equations, equations~(\ref{eq:maxwell1}) and (\ref{eq:maxwell2}), are the familiar
\begin{align}
\dd_t \B &= - \curlE , \nonumber \\
\dd_t \E &= \curlB - \J,
\label{eq:maxwell}
\end{align}
using Heaviside-Lorentz units with $c=1$; this just means that our current density is $4\pi$ times the current density in Gaussian units. In MHD, an additional relation must be given for the current, closing the equations. In force-free electrodynamics, the current is uniquely determined by equations~(\ref{eq:forcefree}) and (\ref{eq:maxwell}), together with the condition $\dd_t \lp \E\cdot\B \rp = 0$, to be \citep{Gruzinov:1999p3173}
\beq
\J = \frac{\vec{B} \cdot \curlB - \vec{E} \cdot \curlE }{B^2} \B + \divE \, \frac{\vec{E}\times\vec{B}}{B^2}.
\label{eq:current}
\eeq
Ohm's law in force-free electrodynamics is therefore essentially geometrical. The first term in equation~(\ref{eq:current}), is the conduction current parallel to $\B$, which maintains the degeneracy condition, equation~(\ref{eq:edotb}). The second term is the drift current, being in the form $\rhoe \vec{v}_{\rm drift}$, where
\beq
\vec{v}_{\rm drift} = \frac{\vec{E}\times\vec{B}}{B^2}
\eeq
is the velocity of the magnetic field lines. It is apparent that there is a second condition,
\beq
B^2 - E^2 > 0,
\label{eq:b2e2}
\eeq
equivalent to requiring the drift velocity to be less than the speed of light; since charged particles cannot cross field lines, this is a requirement if we assume that a macroscopic matter velocity field exists. Equations~(\ref{eq:edotb}) and (\ref{eq:b2e2}) are commonly referred to as the `force-free conditions.' The second condition implies that the electromagnetic field is intrinsically magnetic, in that a frame exists in which the electric field vanishes. 

It is possible for the fields to evolve from a configuration in which this second force-free condition is satisfied everywhere to one in which it is violated at some point, line, surface, or volume. This local breakdown of the force-free approximation is necessarily accompanied by dissipation, as degeneracy is broken and $\E\cdot\B \neq 0$. In general, any configuration having field lines of different topology, such as the open and closed lines of the pulsar magnetosphere, will have points, lines, or surfaces at which $|\B| = 0$, violating the second condition \citep{Uchida:1997p4042}. These sites of force-free breakdown are especially interesting, as the localised dissipation may be responsible for observed radiation.

Force-free electrodynamics supports two classes of waves, which we describe in a frame in which $\E = 0$ \citep{2003ApJ...583..842P}. Fast waves, equivalent to vacuum electromagnetic waves, propagate isotropically with both phase and group velocities equal to the speed of light. They do not carry any charges or currents. Alfv\'{e}n waves can carry charges and currents, have phase velocity $v_{\rm phase} = \pm c \cos\theta$, where $\theta$ is the angle between $\B$ and the wave vector, and have group velocity equal to $c$ and directed along $\B$, $\vec{v}_{\rm group} = \pm c \B/B$.

The `field-evolution' approach we take is not the only way to write the evolutionary equations of force-free electrodynamics. One could equivalently evolve the drift velocity or the Poynting flux vector, $\vec{S} = \E\times\B$, instead of the electric field. The equations can also be written, using Euler potentials, as a Hamiltonian system \citep{Uchida:1997p4042}, or in an axionic formulation \citep{Thompson:1998p4118}.

\section{Numerical method}
\label{sec:nummethod}
\subsection{ The pseudospectral method }

A function $u(x, t)$, the solution of a time-dependent partial differential equation, can be expanded in terms of a set of orthogonal spatial basis functions $\phi_k(x)$,
\begin{equation}
u_N(x, t) = \sum_{k=0}^{N-1} a_k(t) \phi_k(x),
\label{eq:u_N}
\end{equation} 
where $a_k(t)$ are time-dependent expansion coefficients; $u_N$ is an approximation to the function $u$ for some choice of basis functions and truncation $N$. Spatial derivatives can be taken by analytically differentiating $u_N$, since the exact derivatives of the basis functions are known. Considering first an equation in $x$ only,
$$
D u(x) = f(x),
$$
where $D$ is a general differential operator and $f$ is a forcing function, we can think of solving this equation by minimising the residual $R$: $R = D u_N - f$. In what sense we choose to minimise $R$ will determine the kind of spectral method we construct. In the Galerkin (sometimes just called `the spectral') method, the residual is made orthogonal to the basis functions: $(\phi_k, R)=0, \; k=0,\ldots,N-1$, where the brackets indicate an inner product, $(f, g) \equiv \int \omega(x) f(x) g(x) dx$, over a weight function $\omega(x)$.  Since the first $N$ spectral coefficients are exact, $u_N$ can be considered to be a \textit{truncation} of the infinite series expansion.

In the pseudospectral method, the residual is made zero at a set of `collocation' points, $\{x_i\}$: $(\delta[x-x_i], R) = 0,\; i=0, \ldots, N-1$, where $\delta(x)$ is the Dirac delta-function. The resulting $u_N$ is then an \textit{interpolant} of the true $u$, at the chosen grid points. It can be shown that, if the collocation points are chosen as the abscissas of a Gaussian quadrature associated with the basis set and this quadrature rule is used to calculate the inner products, then the Galerkin and pseudospectral methods are equivalent for linear problems; the error penalty for choosing interpolation over truncation is at worst a factor of two, for trigonometric functions and Chebyshev polynomials \citep{Boyd:2001}.

Gaussian quadrature of a function $f$ over a weight function $\omega$,
\begin{equation}
\int_a^b f(x) \omega(x) dx = \sum_{i=0}^{N-1} w_i f(x_i),
\end{equation}
is accomplished by finding the corresponding set of $N$ weights $\{w_i\}$ and $N$ interpolation points $\{x_i\}$; the pay off for being restricted to these interpolation points is that the resulting formula is exact for all $f(x)$ which are polynomials of degree $2N-1$ or less. The weight function determines the basis functions; for example, the Chebyshev polynomials are those polynomials which are orthogonal with respect to the weight $\omega(x) = 1/\sqrt{1 - x^2}$ on the interval $[-1,1]$. For periodic $f(x)$, the composite-trapezoidal and midpoint rules are Gaussian quadratures with an equispaced grid, the corresponding basis being trigonometric functions.

The pseudospectral method can also be thought of as the limiting case of increasing-order finite-difference methods, where the derivative stencil now extends over all grid points. In particular, the Fourier pseudospectral method on a periodic uniform grid is recovered by a finite-difference formalism as stencil width (and formal order of accuracy) goes to infinity \citep{Fornberg:1996}. This approach gives a dense differentiation matrix, whose application requires $O(N^2)$ operations. However, identical derivatives can be calculated for the Fourier and Chebyshev basis sets by using the fast Fourier transform (FFT), which requires only $O(N\ln{N})$ operations; this is sometimes referred to as the `transform method' \citep{Orszag:1970p2427}. In this case, a forward FFT gives the expansion coefficients, $\{a_k\}$, from which can be constructed the coefficients for the derivative series $\{a'_k\}$: $u'_N (x) = \sum_{k=0}^N a'_k \phi_k(x)$. In the Fourier basis, $\phi_k(x) = e^{i k x}$ and so $a'_k = i k a_k$; with Chebyshev polynomials a three-term recurrence relation is used. Finally the derivative at the grid points, $u'_N$, can be found with an inverse FFT. Given a function to be differentiated, this procedure can be thought of as finding an interpolating function of order $N$, at a set of $N$ points, and taking the exact analytic derivative of this interpolating function.

The great benefit of these methods is that spectral approximation is exponentially convergent for sufficiently smooth functions: the error decreases faster than any power of the truncation $N$. It has generally been found that this carries over into exponential convergence of spectral solutions of PDEs, even those with fixed-order time marching. This accuracy has made spectral methods popular in many areas of physics, including meteorology, seismology, shock waves, and reactive flows. Astrophysical applications include accretion disc magnetohydrodynamics \citep*{Chan:2005p616,Chan:2009p607} and general relativity \citep[e.g.][]{1991A&A...252..651G, Bonazzola:1999, Kidder:2000p603, Dimmelmeier:2005p4003, Grandclement:2009}. In engineering electrodynamics, the `pseudospectral time-domain' method was introduced by \citet{Liu:1997p821}, where it was shown to have much lower diffusion and dispersion error than finite-difference methods, and to require either two (Fourier) or $\pi$ (Chebyshev) points per wavelength for adequate resolution, in comparison to eight to sixteen points for finite differences. These lower required grid densities make a spectral method more efficient for achieving a given accuracy, despite the higher number of operations per grid point.

\subsection{Spatial discretisation}
\label{sec:discrete}
In order to simplify eventual extension to curved space-time, we adopt a scheme which allows the use of an arbitrary spatial metric. We store, and advance in time, the contravariant vector components of $\vec{B}$ and $\vec{E}$; curls are taken with $(\nabla\times\vec{F})^i = (1/\sqrt{\gamma}) e^{ijk} \dd_j F_k$ and divergences with $\nabla\cdot\vec{F} =   (1/\sqrt{\gamma})\dd_i(\sqrt{\gamma} F^i)$, where $\gamma$ is the spatial metric determinant, $e^{ijk}$ the Levi-Civita symbol, and $\vec{F}$ stands for $\B$ or $\E$. The quantity to be differentiated, either $F_k$ or $\sqrt{\gamma}F^k$, is first calculated at each point from the contravariant components $F^k$ and the metric, then expanded in orthogonal basis functions. This method requires more forward transforms than one where the derivatives are simplified by the chain rule, since for example both $F_r$ and $\sqrt{\gamma}F^r$ must be transformed into spectral space, but we find it to be more accurate.

Our grid is defined in axisymmetric spherical coordinates $(r, \theta)$, with $N$ collocation points in $r$ and $L$ points in $\theta$. We will use $i$ and $j$ to index grid points, and $n$ and $l$ to index wavenumbers, along each direction; $i, n = 0, \ldots, N-1$, and $j, l = 0, \ldots, L-1$.

In the radial direction we use Chebyshev polynomials, $T_n$, and the collocation points are chosen to be the Chebyshev-Gauss-Lobatto nodes, 
\beq
x_i = - \cos\left( \frac{\pi i}{N-1} \right), \qquad i = 0, \ldots, N-1, \qquad x \in [-1, 1]\;.
\eeq
These can be mapped directly onto the physical coordinate, $r_i = \rin + (\rout - \rin)(1 + x_i)/2$, $r \in [\rin, \rout]$, or via an additional coordinate mapping (section \ref{sec:map}). Since the Chebyshev polynomials are mapped cosine functions,
\beq
T_n(\cos[q]) = \cos(n q)\;,
\eeq
with this choice of grid the Chebyshev transform of a function $f$ can be performed with a fast cosine transform:
\begin{align}
f(x_i) &= \sum_{n=0}^{N-1} f_n T_n(x_i) \;,\nonumber\\
         &= \sum_{n=0}^{N-1} f_n \cos\left( \frac{\pi n i}{N-1} \right).
\end{align} 

In the meridional direction we expand in sine or cosine functions, depending on whether the vector component in question is an even or odd function across the pole (its parity). This is related to the `double Fourier' method of expanding functions on a sphere \citep{Merilees:1974p2212, Orszag:1974p2624}, which is attractive because it avoids the slow Legendre transform required by spherical harmonics. The following are even, and can be expanded in cosines: $F_r, F^r, F_{\phi}, F^{\phi}, \sqrt{\gamma} F^{\theta}$, whereas odd functions that can be expanded in sines are $F_{\theta}, F^{\theta}, \sqrt{\gamma} F^r, \sqrt{\gamma} F^{\phi}$. Here we are only concerned with axisymmetric modes; in general the parity will depend on whether the azimuthal wavenumber is even or odd. To avoid solving the equations directly on the poles we use a shifted grid:
\beq
\theta_j = \frac{j + 1/2}{L}\, \pi\,, \qquad j = 0, \ldots, L-1 \;.
\eeq

For instance, the covariant radial component of the magnetic field, once formed by direct index lowering with the metric, can be expanded as
\beq
B_r = \sum_{n=0}^{N-1} \sum_{l=0}^{L-1} a_{nl} T_n(r) \cos(l \theta)\;,
\eeq
and the combination $\sqrt{\gamma} E^r$, required to calculate $\nabla \cdot \vec{E}$, as
\beq
\sqrt{\gamma} E^r = \sum_{n=0}^{N-1} \sum_{l=0}^{L-1} a_{nl} T_n(r) \sin(l \theta)\;.
\eeq

Once the coefficients $a_{nl}$ of a function have been found, by taking the forward transforms in both directions,  the coefficients of the derivative series $a'_{nl}$ can be calculated. For differentiation by $\theta$ this is simple: $a'_{nl} = - l a_{nl}$ for functions expanded in cosines, and $a'_{nl} = l a_{nl}$ for those expanded in sines. Radial differentiation requires the three-term recurrence relation, relating the Chebyshev coefficients of a function to those of its derivative,
\begin{align}
a'_{N-1, l} & = 0 \nonumber\\
a'_{N-2,l} &= 2 (N-1) a_{N-1,l} \nonumber\\
c_{n-1} a'_{n-1,l} &= a'_{n+1,l} + 2 n a_{n,l} \;,
\end{align}
where $c_0  = 2$, and all other $c_n = 1$.

\subsection{Coordinate mappings}
\label{sec:map}
The standard Chebyshev nodes, $x_n$, are not a suitable radial grid for our problems. They are very strongly clustered near the endpoints, leading to a time step restriction for hyperbolic problems that goes like $\Delta t \sim O(1/N^2)$, which makes obtaining high spatial resolution in the rest of the domain unnecessarily expensive. A coordinate map can be used to relieve the endpoint clustering, and also put more nodes in the part of the domain where higher resolution is required. Consider a map from $x$ to a new coordinate $y$:
\begin{align*}
y &= g(x) \\
\frac{\dd}{\dd y} &= \frac{1}{g'}\frac{\dd}{\dd x},
\end{align*}
where $g' = d g/d x$. The derivative coefficients can be calculated as before, and the mapping of the derivative achieved by a multiplication in real space. The arcsine map \citep{Kosloff:1993p418},
\begin{align}
x_{\mathrm{asin}} \equiv g_{\mathrm{asin}}(x)  &=  \frac{\arcsin(\alpha x)}{\arcsin(\alpha)} \nonumber\\
g'_{\mathrm{asin}} &= \frac{\alpha}{\arcsin(\alpha)}\frac{1}{\sqrt{1 - \alpha^2 x^2}}\;,
\end{align}
where $\alpha$ is a constant between zero and one and $x_{\mathrm{asin}} \in [-1,1]$, will create an almost equispaced grid with less clustering at the endpoints; the grid stretching becomes more pronounced as $\alpha$ is increased towards unity. Care must be taken in choosing $\alpha$: since the map is singular at the endpoints, choosing $\alpha$ too large would impair the convergence of the series. The maximum discrepancy between a function\footnote{The complex function $f(z)$ is the analytic continuation of the real function $f(x)$.} $f(z)$ and its truncated series representation $P_N(z)$ is
\begin{equation}
\label{eq:kte-err-1}
\mathrm{max} \,| f(z) - P_N(z)| = c \epsilon\;,
\end{equation}
where $c$ is a constant which depends on $f$ but is independent of $N$, and $\epsilon$ is related to $\alpha$ by
\beq
\epsilon = \left( \frac{1 - \sqrt{1 - \alpha^2}}{\alpha}\right)^N \; \longrightarrow \; \alpha = \mathrm{sech}\left( \frac{|\ln\epsilon|}{N}\right)\;.
\eeq
Therefore choosing $\epsilon$ to be sufficiently small will usually make the singularity harmless \citep{Don:1997p1206, Mead:2003p1235}. Equation~(\ref{eq:kte-err-1}) determines the \textit{largest} pointwise error, which usually appears at, or close to, a boundary; the error can be smaller than $\epsilon$ in the interior of the domain, which we confirm in our tests in Section~\ref{sec:tests}.  For fixed $\epsilon$ the minimum grid spacing only decreases as $O(1/N)$. 

The nearly equispaced grid resulting from the arcsine map is still not perfect for computations in spherical coordinates when $\rout \gg \rin$, because the lines of constant $\theta$ converge towards the centre of the domain, and so too much radial resolution is used where the angular resolution is low, and not enough where it is high. We would prefer a grid where $\Delta r \leq r\, \Delta \theta$ over most of the domain. To construct this we use a combination of the arcsine map and a smooth algebraic stretching:
\beq
x_{\rm alg} \equiv g_{\mathrm{alg}}(x) = Q \frac{1 +  x_{\rm asin}}{Q + 1 - x_{\rm asin}}  - 1 \;,
\eeq
where $Q$ is a constant and $x_{\rm alg} \in [-1,1]$. The grid is then linearly transformed to the desired physical coordinates
\beq
r = R_{\rm in} + \frac{1}{2}(R_{\rm out} - R_{\rm in})(1 + x_{\rm alg}), \quad r \in [R_{\rm in}, R_{\rm out}]\;.
\eeq
We generally use $Q \sim 0.1 - 1$, and set the map-induced error to $\epsilon \sim 10^{-9} - 10^{-15}$. It appears to be preferable to have the radial grid spacing somewhat smaller than the meridional spacing throughout most of the domain. 
Figure~{\ref{fig:radialgrid} shows the inter-nodal spacings of an example grid.

\begin{figure}
\includegraphics[width=84mm]{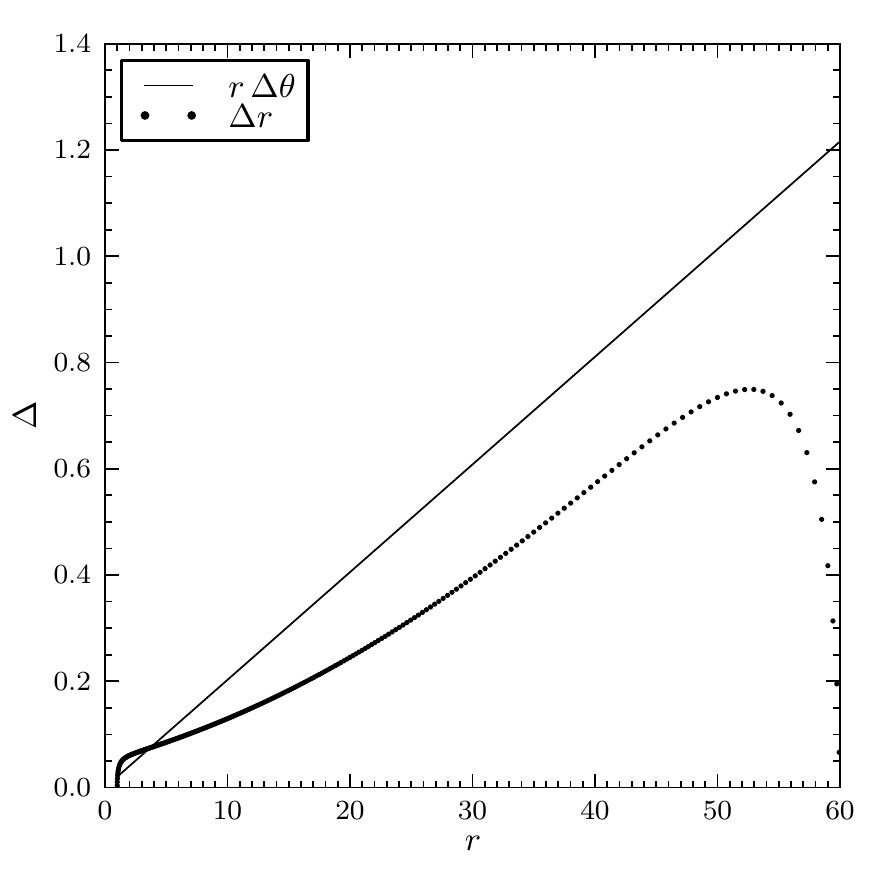}
\caption{\label{fig:radialgrid} Grid spacing versus radius: Chebyshev nodes with a combination arcsine + algebraic mapping; $N=256$, $L=155$, $Q = 0.6$, $\epsilon = 10^{-11}$. }
\end{figure}

Some of our models also include a coordinate transformation in the meridional direction, in order to increase the resolution around a reconnecting current sheet in the equatorial plane. To this end we employ the `Kepler-Burgers' mapping \citep{Boyd:1992p971},
\beq
\tilde{\theta} = \theta + \frac{\gamma}{2}  \sin (2\theta),
\eeq
where $\theta$ is evenly spaced on $[0,\pi]$ and $\tilde{\theta}$ is the new stretched coordinate. The constant $\gamma$ controls the degree of stretching; when using this map we set $\gamma \sim 0.3 - 0.5$.

\subsection{Time evolution}
\label{sec:time}
When the spatial derivatives have been found the system of equations becomes a set of ordinary differential equations in time, one for each vector component, which we solve with an explicit Runge-Kutta integrator. We mainly use a third-order, three stage, method \citep{Fornberg:1996}. If the ODE to be solved is
$$
\frac{d u}{d t} = f(u, t),
$$
$u \in \left\{ B^i, E^i\right\}$, then we can find the solution at time step $n+1$, $u^{n+1}$, from that at time $n$ by
\begin{align}
k^{(1)} &= \Delta t f(u^n, t^n) \notag \\
k^{(2)} &= \Delta t f(u^n + \frac{1}{3} k^{(1)}, t^n + \frac{1}{3} \Delta t) \notag \\
k^{(3)} &= \Delta t f(u^n + \frac{2}{3} k^{(2)}, t^n + \frac{2}{3} \Delta t) \notag \\
u^{n+1} &= u^n + \frac{1}{4}(k^{(1)} + 3 k^{(3)})  \label{eq:rk3} \,.
\end{align}
Since $k^{(2)}$ is not required once $k^{(3)}$ has been calculated, the latter can overwrite the former in memory, and so this method only requires storage for two intermediate arrays. Explicit time advancement is subject to the Courant-Friendrich-Lewey stability constraint on the time step, $\Delta t \leq C_{\rm CFL} \delta_{\rm min}$, where $\delta_{\rm min}$ is the smallest grid spacing, in our case always found at the inner surface. The factor $C_{\rm CFL}$ is, in general, problem and time-integrator dependent; we find that equation~(\ref{eq:rk3}) is stable for $C_{\rm CFL} \leq 1$, and hence always set $\Delta t = \delta_{\rm min}$.

We compared the above with a fifth-order, six stage, Runge-Kutta integrator. The improvement in accuracy is negligible, even for our stringent test with the Michel monopole solution, indicating that time-stepping errors are subdominant. This is unsurprising given that a fine grid in the radial direction is required near the stellar surface to produce highly-accurate solutions, and so the stability-limited time step produces small errors.

\subsection{Spectral filtering}
\label{sec:filtering}
The use of spectral methods for non-linear hyperbolic problems comes with two principal difficulties. The first is the build-up of power at high wavenumbers due to non-linear couplings between lower wavenumbers. Hyperbolic problems have no explicit dissipation in the equations of motion, and spectral methods have very low intrinsic dissipation, so these high modes do not decay, and can lead to the breakdown of the scheme by the aliasing instability. Aliasing occurs because a discrete transform of point values mistakes high-frequency components for low; all modes $e^{i(\omega + qN)x}$, for $q = 0, \pm \pi, \pm 2\pi \ldots$, will be identical when represented on a grid of $N$ points in the interval $[-\pi,\pi]$. Non-linear terms produce high-frequency modes, which are then mistaken as low-frequency modes by the discrete transforms; these phantom low-frequency components are then similarly combined to generate more power at high frequencies, a clearly unstable cycle.

The second difficulty is the ability of non-linear couplings to create discontinuities in a solution which was previously smooth. Spurious oscillations appear in the spectral interpolant when a function is not resolved by the grid\footnote{Any set of points on a grid will be faithfully recovered following transforms into, and back out of, Fourier space; however, the interpolating function, constructed from the Fourier coefficients, can show spurious oscillations \textit{between} the grid points, causing oscillations in the derivative \textit{at} the grid points.}; this is the Gibbs phenomenon. A jump discontinuity causes $O(1)$ errors in its immediate vicinity, and reduces the convergence rate to first order elsewhere.

Both of these difficulties can be largely overcome with spectral filters. If a function $u$ is expanded in a basis $\phi_n$, its filtered approximation, $\mathcal{F}u$, is given by 
\beq
\mathcal{F}u = \sum_{n = 0}^{N-1} \sigma\left( \frac{n}{N-1} \right) \tilde{u}_n\phi_n \;,
\eeq
where $\tilde{u}_n$ are the discrete expansion coefficients. \citet{Vandeven:1991p3486} showed that if the filter function, $\sigma(\eta)$, is unity at $\eta=0$, zero for all $|\eta| \geq 1$, and has at least $2p-1$ continuous derivatives, then $\mathcal{F}u$ will converge with $N$ at $2p$-th order even in the presence of a jump discontinuity, except very close to the jump. In addition, since it strongly damps the high modes, such a filter can prevent the onset of the aliasing instability if regularly applied to each field in a time-dependent simulation. We tried two variable-order examples, the erfc-log filter \citep{Boyd:1996p3494},
\begin{align}
\bar{\eta} &\equiv |\eta| - \frac{1}{2} \notag \\
\sigma_{\rm erfc-log} (\eta) &= \frac{1}{2}\mathrm{erfc}\left\{ 2 (2p)^{1/2} \bar{\eta} \sqrt{ \frac{-\ln (1 - 4 \bar{\eta}^2)}{4 \bar{\eta}^2} } \right\}\;,
\end{align}
and the exponential filter \citep{Majda:1978},
\beq
\sigma_{\rm exp} (\eta) = e^{-\alpha \eta^{2p}}\;.
\label{eq:expfilt}
\eeq
The latter can be made to fulfil approximately the requirement of being zero for all $|\eta| \geq 1$ by setting $\alpha = \alpha_M = -\ln(\epsilon_M)$, where $\epsilon_M$ is machine precision; we use $\alpha_M = 35$. We find the exponential filter to give more accurate results and to allow weaker filtering, and so use it exclusively.

The use of the exponential filter to control aliasing was studied in detail by \citet{Hou:2007p3289}, where they found that a high-order ($2p = 36$) filter can prevent instability in marginally-resolved fluid dynamics simulations while producing more accurate solutions than standard dealiasing methods. We also find this to be the case for our equation system, even though the non-linear coupling is much stronger than in Euler's equations. For all science runs we use a filter with $\alpha = \alpha_M$ and $2p = 36$, which appears to balance well the conflicting demands of assuring stability while minimising unphysical dissipation. When very low numbers of modes are used (roughly $N < 48$ in any direction) the filtering order needs to be reduced somewhat. This high-order filter is applied to the coefficients of every derivative series\footnote{When using Chebyshev polynomials it is important to filter the coefficients of the derivative series, $a'_{nl}$, rather than those of the function before the recurrence relation is used,  the $a_{nl}$ \citep{Godon:1993p724}. For this reason our filtering operations are implemented inside the coefficients-to-grid inverse transform.}, and directly to the coefficients of the solution itself at the end of each full Runge-Kutta time step.

We turn now to the second issue, concerning stability and convergence in the presence of jump discontinuities. Current sheets, regions of formally infinite current density implying discontinuities in the magnetic field, are a generic feature of force-free electrodynamics. The danger is that the strong production of high-wavenumber power, or the non-linear interaction of the Gibbs oscillations with the solution, will lead to instability, or at least loss of convergence.  Dissipation is required to prevent this, and ensure that the correct entropy solution is selected; see \citet{Gottlieb:2001p348} for an extensive review of stability and convergence theory for non-linear hyperbolic problems.

This dissipation can be effectively, and efficiently, provided by a spectral filter \citep{Tadmor:1993, Don:1994p1199}. For a time-dependent problem,
$$
\frac{d u}{d t} = f(u)\;,
$$
we can advance the solution one time step, and then apply a filter to the solution, $u(t + \Delta t)$.  Note that in this case we are using the term `filter' more broadly than above; in particular, we will not require that $\sigma(1)=0$. If we choose to use the exponential filter, equation~(\ref{eq:expfilt}),  then the modified equation, taking into account the action of the filter, is
\beq
\frac{\dd u}{\dd t} = f(u) - \alpha \frac{(-1)^p}{\Delta t\, N^{2p}} \frac{\dd^{2p} u}{\dd x^{2p}} + O(\Delta t^2) \;,
\eeq
if $u$ is expanded in Fourier series, and
\beq
\frac{\dd u}{\dd t} = f(u) - \alpha \frac{(-1)^p}{\Delta t\, N^{2p}} \left[ \sqrt{1-x^2} \frac{\dd}{\dd x}  \right]^{2p} u + O(\Delta t^2) \;,
\eeq
if it is expanded in Chebyshev polynomials. The second result follows from the relation
\beq
\left[ \sqrt{1-x^2} \frac{\dd}{\dd x} \right]^2 T_n(x) + n^2 T_n(x) = 0\;.
\eeq
In this case the dissipation decreases to zero at the boundaries (recall $x \in [-1,1]$, and here $u(x)$ is non-periodic), which is useful since then no additional boundary conditions are required \citep{Boyd:1998p3611}.

Applying an exponential filter is therefore similar to adding a hyperviscous term to the equation, where the magnitude of the hyperviscosity is
\beq
\epsilon_N = \frac{\alpha}{\Delta t\, N^{2p}}\;.
\label{eq:epsn1}
\eeq
The action of the filter is equivalent to an implicit time integration of the hyperviscous term, and so no additional stiffness is added to the equations. 

The similarity of the effect of the exponential filter to an explicit hyperviscous term allows us to import stability and convergence theory derived using such terms. The spectral viscosity (SV) method \citep{Tadmor:1989p237, Tadmor:1990p3462} uses second-order viscous regularisation ($2p = 2$), and convergence is obtained by excluding an increasing fraction of low-wavenumber modes from the viscous term as $N$ is increased. Extensions to higher order, $2p \geq 4$, followed for schemes based on Fourier \citep{Tadmor:1993} and Chebyshev \citep{Ma:1998p321} expansions, known as the super spectral viscosity (SSV) methods. These can be proven to converge to the correct solution of a scalar conservation law for
\begin{align}
\epsilon_N &= \frac{C}{N^{2p-1}}\;, \label{eq:epsofN}
\\
p &\leq O(\ln N) \notag \;. 
\label{eq:hviscscaling}
\end{align}
Convergence can't be guaranteed for a system of equations, although it has been proven that if the scheme converges, it converges to the correct entropy solution \citep{Carpenter:2003}. Despite the lack of solid theoretical results for non-linear systems, experience has shown that the method can be stable and convergent; multidimensional examples include shock-vortex interaction \citep{Don:1994p1199, Sun:2006}, and problems involving both shocks and combustion \citep{Don:1998p188, Gottlieb:2005p2373}, where spectral methods were found to perform well in comparison with high-order shock-capturing schemes.

Using equation~(\ref{eq:epsofN}) as a guide, we find from equation~(\ref{eq:epsn1}) that the filter amplitude should scale as $\alpha = \alpha_{\rm SSV} = C N \Delta t$. The time step scales as $\Delta t \propto 1/N$ (Section~\ref{sec:map}), and so $\alpha_{\rm SSV}$ should be roughly constant. The filter order, $2p$, should only increase slowly with $N$, and so we set it to a constant as well. Numerical experiments confirm that fixed $\alpha_{\rm SSV}$ and $p$, determined by low-resolution simulations, lead to stable and convergent results as resolution is increased. We find best results are obtained with $2p = 8$, $\alpha_{\rm SSV} \sim 0.01 - 0.1$, corresponding to a weak hyperdiffusion which decreases in strength with resolution like $N^{-7}$. The SSV filters are applied to every component of the electric and magnetic fields at the end of each full Runge-Kutta time step.

To summarise the filtering procedure, we apply a very high order filter, with $\alpha=35$ and $2p \sim 36$, to the inverse transform of every derivative series. At the end of each full time step, we apply both the previous filter and one with $\alpha \sim 0.01 - 0.1$ and $2p=8$ to the field variables themselves.

\subsection{Post-processing}
\label{sec:postprocess}
The SSV method is exponentially convergent in any error norm, if $p$ increases linearly with $N$; however stability will often not permit this, and in any case $O(1)$ errors will remain near any discontinuities. It has long been argued that high-order methods retain enough information to reconstruct a highly accurate (and sometimes spectrally convergent) solution, even if oscillations are present and the pointwise errors are large. The Gibbs oscillations are not noise, and they can be safely removed, without destroying the accuracy of the underlying solution, with a post-processing step after the simulation has finished. There are many ways to do this: for example, real-space filtering or mollification \citep{Gottlieb:1985}, one-sided filters \citep*{Cai:1992p3833}, spatially-varying spectral filters \citep{Boyd:1996p3494, Tadmor:2005}, and reprojection into a Gibbs-complementary set of basis functions \citep{Gottlieb:1992p3754, Gelb:2006p3837}. The reprojection method is particularly popular, and many examples exist of successful one-dimensional  reconstructions of oscillation-free solutions \citep[e.g.][]{Shu:1995p3488, Sarra:2003p352, Ma:2006p197}.

The problem with all of the above methods is that they require the locations of the discontinuities to be known accurately, which is particularly difficult in more than one dimension. We were unable to find a sufficiently robust means of determining the number of unresolved features and their locations, given that the physics generates, and our scheme is capable of resolving, oscillations with wavelengths close to the grid scale. 

We show here an example of post-processing applied to a solution with a known discontinuity---the equatorial current sheet in the aligned rotator problem (Section~\ref{sec:pulsar}). Figure~\ref{fig:postproc} shows (a) the original SSV solution before post-processing, (b) after applying the optimal spatially-varying spectral filter \citep{Tanner:2006p2010}:
\begin{align}
& \sigma_{\rm opt} (k, N, x) = e^{-z} \sum_{n=0}^{\lfloor \kappa N d(x) \rfloor} \frac{1}{n!}  z^n   ,\\
& z = \frac{\alpha k^2 d(x)}{2 N} \nonumber
\end{align}
where $d(x)$ is the distance to the nearest discontinuity and $\kappa$ and $\alpha$ are constants, and finally (c) using the digital total variation (DTV) spatial filter \citep{Rudin:1992p2148, Sonar:2002p2138}. The DTV filtered solution $u$ to a noisy variable $u^0$ is found by minimising the fitted total variation energy
\begin{align}
& W_{\rm DTV} = \sum_{\beta} |\nabla u_{\beta}| + \frac{\lambda}{2}\lp u_{\beta} - u^0_{\beta} \rp^2, \\
&  |\nabla u_{\beta}| = \sqrt{\sum_{\gamma} (u_{\beta} - u_{\gamma})^2} \nonumber,
\end{align}
where $\beta$ ranges over all points in the dataset, $\gamma$ denotes each point's neighbours, and $\lambda$ is related to the expected noise level. The minimisation can be done by linearised Jacobi iteration. This method does not require the locations of the discontinuities, has a natural multi-dimensional form, requires only an estimate of the size of the oscillations to be removed, and has been applied to Chebyshev-based spectral methods with good results \citep{Sarra:2006p2103}. However, we find it to perform poorly if resolved physical high-frequency oscillations are present, even with an adaptive local noise estimate.

\begin{figure}
\subfigure[No post-processing]{ \includegraphics[width=8cm]{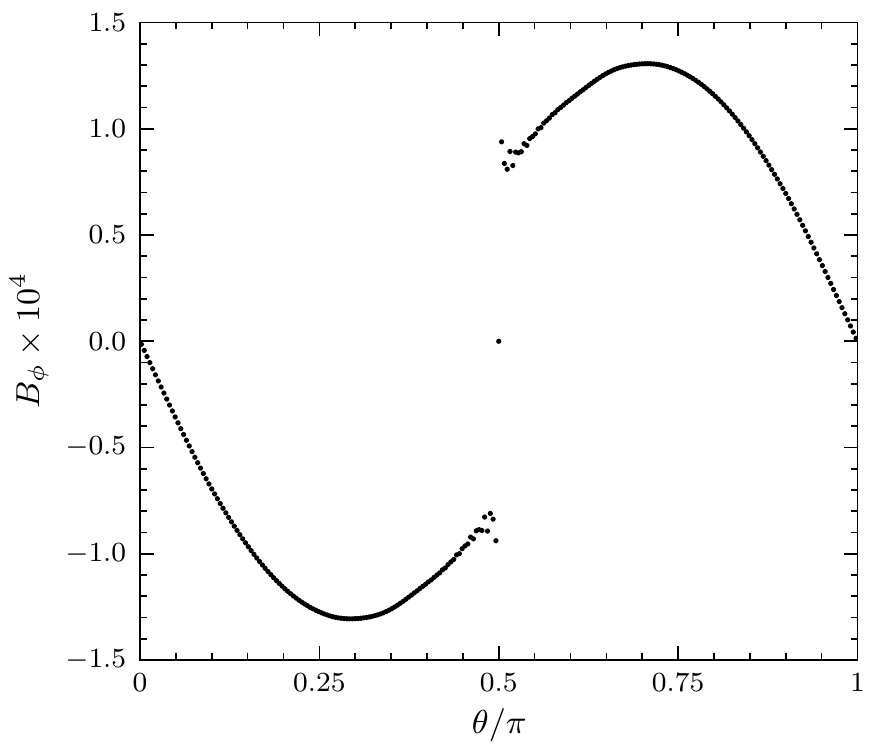} }
\subfigure[Optimal filter]{ \includegraphics[width=8cm]{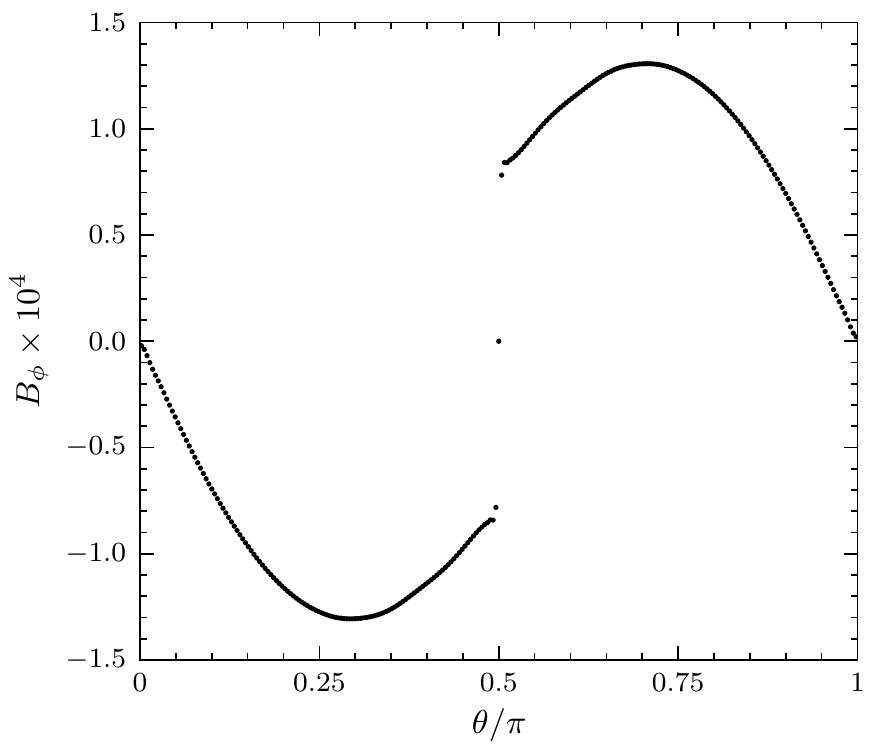} }
\subfigure[Digital total variation]{ \includegraphics[width=8cm]{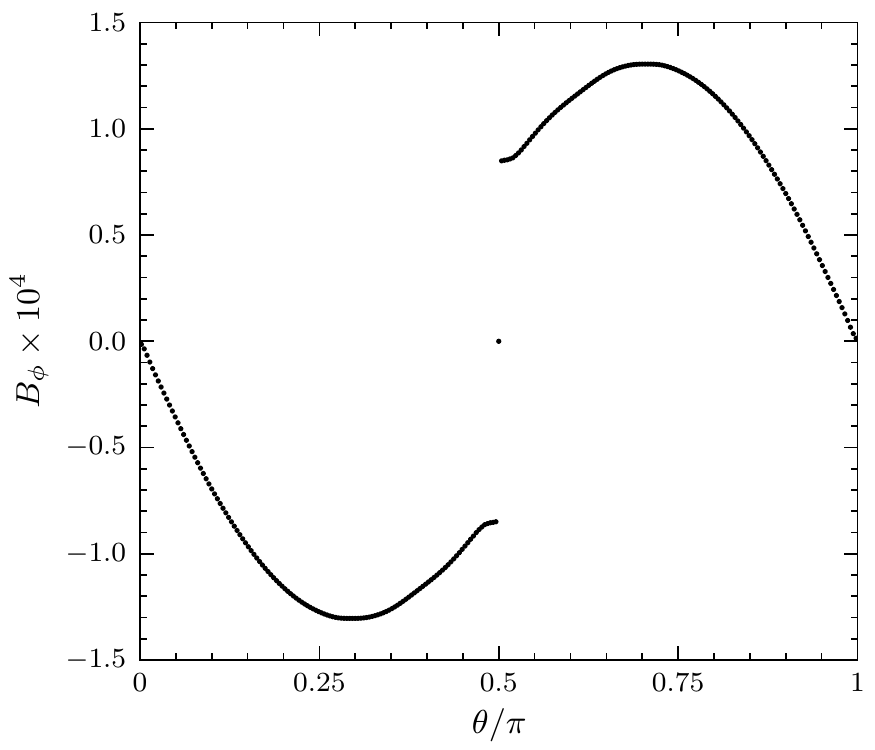} }
\caption{\label{fig:postproc} Toroidal magnetic field with discontinuity (current sheet), $L=255$: (a) no post-processing, SSV only; (b) with optimal filter, $\kappa=0.05$, $\alpha=1$; (c) with DTV reconstruction.}
\end{figure}

We do not use any post-processing technique for our results in Sections~\ref{sec:tests} and \ref{sec:pulsar}, with the exception of Fig.~\ref{fig:chargecontour} where we apply a two-dimensional DTV filter. Our intention is simply to highlight that any Gibbs oscillations which are present do not destroy the accuracy of the scheme. The pointwise errors may be large near a jump, and we do not claim spectral convergence of the field values on the grid, but the overall evolution of the system is still consistent with high-order accuracy.

\subsection{Force-free current}
\label{sec:current}
The current density function required to close Maxwell's equations with force-free dynamics can be written as
\begin{align}\label{eq:J2}
\vec{J} &= \vec{J}_{\parallel} + \vec{J}_{\rm drift}, \nonumber\\
\vec{J}_{\parallel} &= \frac{\vec{B} \cdot \curlB - \vec{E} \cdot \curlE }{B^2} \B, 
\; \qquad
\vec{J}_{\rm drift} = \divE \, \frac{\vec{E}\times\vec{B}}{B^2}.
\end{align}
The effect of $\J_{\parallel}$ is to maintain the force-free condition $\E \cdot \B = 0$. To date, schemes that rely on staggered grids do not explicitly include this term, because it demands that all components of the fields and their curls be evaluated, by interpolation if necessary, at the locations where the electric field components are defined. Instead, the effect of the parallel current is mimicked by resetting the values of $\E$ on the grid such that $\E_{\parallel} = 0$. Since the inherent accuracy of spectral derivatives frees us from the need for staggered grids, we can evaluate the full current function without interpolation, and so include both $\J_{\parallel}$ and $\J_{\rm drift}$ in the equations of motion. We also manually set $\E_{\parallel}$ to zero at the end of each full time step; this has essentially no discernible effect in most of the domain since $\J_{\parallel}$ keeps $\E \cdot \B$ very close to zero anyway, but gives slightly cleaner evolution close to the inner boundary at lower grid resolution. Specifically, $\E$ is projected, parallel to $\B$, into a plane perpendicular to $\B$:
\beq
\E \rightarrow \E - (\E\cdot\B)\frac{\B}{B^2}.
\label{eq:eproject}
\eeq
Retention of the parallel current results in much lower diffusion error; see the twisted dipole test in Section~\ref{sec:twist}.

Force-free evolution can lead to configurations in which the second force-free condition, $B^2 - E^2 > 0$, is violated. For example, in the equatorial current sheet of the axisymmetric rotating dipole (Section~\ref{sec:pulsar}) all components of the magnetic field are zero, and any electric field results in the violation of the condition. An ideal force-free configuration requires $\E=0$ in the current sheet. This can be simulated by immediately reducing the magnitude of the electric field if it is larger than the magnitude of the magnetic field, such that $B^2 - E^2 = 0$ at the end of the operation. At each Runge-Kutta substep, if the condition is violated, we shrink the electric field vector, leaving its direction unchanged:
\beq
\E \rightarrow \sqrt{\frac{B^2}{E^2}} \, \E .
\eeq
This removal of electric field acts like a small, highly localised, source of dissipation in current sheets. Physically, the removed electromagnetic energy would be converted to thermal energy and radiation; in these simulations it is simply lost from the system.

\subsection{Boundary conditions}

A system of hyperbolic equations generally requires suitable conditions to be provided at the boundaries of the computational domain. These boundary conditions will depend on the spatial geometry and the physical problem under investigation. We often wish to simulate a volume of force-free plasma surrounding a neutron star, and so describe the boundary treatment for this case.

In these simulations, the inner boundary, $r = \rin$, corresponds to the stellar surface, while the outer boundary at $\rout$ should as closely as possible behave as a membrane which perfectly transmits outgoing waves without generating any incoming waves. The behavioural boundary conditions at the poles are satisfied automatically by the choice of either sines or cosines as the basis functions in the meridional direction, as described in Section~\ref{sec:discrete}.

\subsubsection{Inner boundary}

For our purposes the star is a perfect rigid conductor, all or part of which can be rotated. Since the star is a perfect conductor, the electric field will be zero in the rotating frame, $\vec{E}' = 0$. The fields in the lab frame are given by
\begin{align}
\B &= \B', \nonumber\\
\E &= - \lp \vec{\Omega}\times\vec{r} \rp \times \B',
\end{align}
where $\vec{\Omega}$ is the local angular velocity vector, and therefore
\beq
\E = - \lp \vec{\Omega}\times\vec{r} \rp \times \B.
\label{eq:perfconduct}
\eeq 
Rotation about the $z$-axis corresponds to the application of an induced poloidal electric field. 

Equation~(\ref{eq:perfconduct}) provides the relationship between $\E$ and $\B$ infinitesimally below the surface, but our first grid points are in the force-free plasma infinitesimally above. The normal component of the magnetic field, $B^r$, and the tangential components of the electric field are continuous across the surface, and therefore are known. The required boundary values are $B^r = B^r (\theta)$, $E^{\theta} = - \Omega B^r \sin \theta$, and $E^{\phi} = 0$; these are strongly enforced at every Runge-Kutta substep. The other components must be allowed to evolve undisturbed, since they depend on unprescribed surface currents and charges on the star.

A complication introduced by the combination of the above boundary conditions and the SSV filtering of the fields is the anomalous leakage of energy into the domain through the inner boundary. Consider the $B^r$ field: at the end of each time step it is filtered, or smoothed, leading to a slight broadening of the $B^r(r)$ profile, and hence diffusion of field from the boundary into the domain. According to the above boundary condition, stating that $B^r$ is constant in time, this field at the boundary is immediately replenished, and so over time this cycle can increase the volume-integrated energy on the grid. 

A solution is to subtract the initial vacuum fields $(\B_0, \E_0)$ from the dynamical fields $(\B,\E)$, and evolve $\tilde{\B} = \B - \B_0$ and $\tilde{\E} = \E - \E_0$. The current density and electric field in the initial configuration are zero, and so, writing $\B = \B_0 + \tilde{\B}$, $\E = \tilde{\E}$, we get the equations
\begin{align}
\dd_t \tilde{\B} &= - \nabla \times \tilde{\E} \nonumber \\
\dd_t \tilde{\E} &= \nabla \times \tilde{\B} - \J (\B_0 + \tilde{\B}, \tilde{\E}).
\end{align}
Note also that $\nabla \times \B_0 = 0$, and so $\curlB = \nabla \times \tilde{\B}$ in the current density function---no additional derivatives are required. There is no requirement that these new variables be small, and in much of the magnetosphere they will be larger than the corresponding initial field. The initial field is never differentiated or filtered, there is no field leakage by the above mechanism since $\tilde{B}^r = 0$ is the new magnetic boundary condition, and the energy conservation of the code is improved dramatically (see Section~\ref{sec:twist}). 

To reduce notational clutter, these variables will be neither distinguished nor discussed elsewhere in this paper; the only addition they require is that the evolved magnetic field be temporarily added to a stored initial field before being passed to the current density function (which takes as arguments the electric and magnetic fields, and their curls).

Finally, a small, slowly growing, anomalous toroidal magnetic field was found in certain simulations using a dipole base field, in the equatorial region immediately next to the star. This field appeared only when that region was `dead' ($\curlB = 0$). We have attributed this feature to Alfv\'{e}n waves on under-resolved field lines, propagated with a numerical scheme with very low diffusivity. 

This phenomenon is completely eradicated by setting a small circular region of the poloidal plane, centred on $(r, \theta) = (\rin, \pi/2)$ and with radius $r_{\rm  drift}$, to only use the drift current contribution to $\J$ in the equations of motion. Neglecting $J_{\parallel}$ makes the scheme in the `drift' region sufficiently diffusive that the anomalous feature does not develop. Inside this region the electric field is projected to be perpendicular to the magnetic field, using equation~(\ref{eq:eproject}), at the end of each full time step, exactly as in the rest of the domain. The radius of the drift region can be decreased with increasing resolution; we use $r_{\rm drift} \sim$~0.25~$\rin$. We have found that modifying the scheme in this small fraction of the domain does not affect the solution in the rest of the domain.

\subsubsection{Outer boundary}
\label{sec:outerboundary}
We implement the non-reflecting boundary condition at the outer boundary using an approximate characteristic decomposition. Spectral methods are very sensitive to `incorrect' boundary conditions, and the simple zero-gradient condition that works well in low-order methods leads to instability. Characteristic-based boundary treatments \citep[e.g.][]{Abarbanel:1991p762, Godon:1996p720} specify the outgoing characteristic variables using the calculated data on the grid, and set the incoming characteristic variables to zero.

If we construct a six-component vector of the fields like $\vec{q} = \{ \B, \E \}$, then the one-dimensional equations of motion in Cartesian coordinates can be written in the general form
\beq
\dd_t \vec{q} + \mtx{A} \dd_x \vec{q} = 0,
\eeq
where the matrix $\mtx{A}$ (the flux Jacobian) should include contributions from the linear and non-linear terms. This matrix can be decomposed into its eigenvalues and eigenvectors,
\beq
\mtx{A} = \mtx{S} \mtx{\Lambda} \mtx{S}^{-1},
\eeq
$\mtx{\Lambda}$ being a diagonal matrix of the eigenvalues, $\mtx{S}$ containing the right eigenvectors, and $\mtx{S}^{-1}$ the left eigenvectors. The characteristic variables, $\vec{w}$, are found using the left eigenvectors,
\beq
\vec{w} = \mtx{S}^{-1} \vec{q},
\eeq
using which the equations of motion decouple, since $\mtx{\Lambda}$ is diagonal:
\beq
\dd_t \vec{w} + \mtx{\Lambda} \dd_x \vec{w} = 0.
\eeq
In this one-dimensional case each component of $\vec{w}$ will move in either the positive or negative $x$ direction, depending on the sign of the corresponding eigenvalue. The incoming variables can be identified, set to zero, and the primitive variables recovered using $\vec{q} = \mtx{S} \vec{w}$.

Rather than using the exact characteristic variables, we have implemented an approximate boundary condition using the characteristics of the vacuum Maxwell's equations. 

At each point on the outer boundary, construct a local Cartesian vector basis, with $\vec{\hat{x}}$ in the radial direction, $\vec{\hat{y}}$ along $\vec{\hat{\theta}}$, and $\vec{\hat{z}}$ along $\vec{\hat{\phi}}$. We will then identify $B_y = r B^{\theta}$, $B_z = r \sin\theta B^{\phi}$ etc., where the Cartesian components are in a normalised basis, and the spherical components are contravariant as usual. The four propagating characteristic variables are 
\beq \label{eq:vaccvs}
\begin{pmatrix}
w_1 \\
w_2 \\
w_3 \\
w_4
\end{pmatrix}
= \frac{1}{2}
\begin{pmatrix}
E_z + B_y \\
E_y - B_z \\
E_z - B_y \\
E_y + B_z
\end{pmatrix}.
\eeq
The variables $w_1$ and $w_2$ propagate in the negative $r$ direction, these are incoming variables and will be set to zero; $w_3$ and $w_4$ propagate in the positive $r$ direction and will be calculated from the field values on the grid. Inverting equation~(\ref{eq:vaccvs}), zeroing the incoming characteristics, and replacing Cartesian with spherical components gives
\beq
\begin{pmatrix}
B^{\theta} \\
B^{\phi} \\
E^{\theta} \\
E^{\phi}
\end{pmatrix}
=
\frac{1}{r}
\begin{pmatrix}
- w_3 \\
w_4/\sin\theta \\
w_4 \\
w_3/\sin\theta
\end{pmatrix};
\eeq
the indicated fields at the boundary are replaced with these values at every Runge-Kutta substep, the radial components being left unchanged. 

This approximate boundary condition is stable and works well for predominantly radial waves, but can generate artefacts when sizeable tangential waves are present. To prevent these from appearing we use a thin sponge layer next to the outer boundary, which absorbs outgoing waves \citep{If:1987p713}. Any waves reaching the boundary are much attenuated, and so the vacuum boundary condition is a better approximation. Likewise, the approximate non-reflecting condition allows the use of a thinner and weaker absorbing layer than would be sufficient with a reflecting boundary.

The sponge layer is introduced by adding a frictional term to Maxwell's equations, which become
\begin{align}
\dd_t \B &= - \curlE - \sigma_{s}(r) \B_{\rm ang}  \nonumber \\
\dd_t \E &= \curlB - \J - \sigma_{s}(r) \E ,
\end{align}
where $\B_{\rm ang} = \{0, B^{\theta}, B^{\phi} \}$, not including $B^r$---we found that damping the radial magnetic field led to unphysical currents leaking back into the normal domain. This approach is similar to the matched-layer method of \citet{Yang:1997p3936}. The frictional coefficient, $\sigma_s$, is chosen to be zero in most of the domain, and to rise smoothly near the boundary; the functional form we use is 
\beq
\sigma_s(r) = \left \{
\begin{aligned}
&0    &\textrm{ if } r < r_s \\
&\sigma_0 \left( 1 - \exp\left[-\gamma \left( \frac{r - r_s}{\rout - r_s}\right)^\beta\right] \right)   &\textrm{ if } r > r_s
\end{aligned}
\right.
\eeq
with the values $\sigma_0 \sim 0.5 - 1$, $\gamma \sim 6$, $\beta \sim 4$. This boundary treatment is robust, effective, and insensitive to the values of the sponge layer coefficients.

\subsection{Magnetic field divergence}

Maxwell's equations comprise two evolution equations, for the electric and magnetic field vectors, and two constraint equations. One of these constraints, Gauss's law, is automatically satisfied, since the charge density has been replaced by $\divE$ in the drift current term in equation~(\ref{eq:J2}). The other constraint is the solenoidal condition on the magnetic field: $\divB = 0$. 

In theory this should not be a concern, since the evolution equation for $\B$ implies $\dd_t (\divB) = - \nabla\cdot(\curlE) = 0$. However, in numerical schemes the operators for calculating divergences and curls usually do not satisfy $\nabla\cdot(\nabla\times\vec{V}) = 0$ exactly for any vector $\vec{V}$, raising the worry that this truncation-level magnetic field divergence might build up over time. The presence of such magnetic charges can lead to unphysical forces and even instability.

To maintain stability many MHD codes use constrained transport, which can ensure that some representation of $\divB$ is kept at machine zero (see \citet{Toth:2000} for a review). These methods are incompatible with our global spatial derivatives. 

Another option is to evolve the magnetic field using a vector potential: writing $\B = \nabla\times\vec{A}$, the first evolution equation becomes $\dd_t\vec{A} = - \E$; see \citet{Chan:2009p607} for a spectral implementation. Although $\nabla\cdot(\nabla\times\vec{A})$ is only zero to truncation error, this error would not grow over time. The disadvantage of a vector potential is the introduction of second-order spatial derivatives, and, in our case, an increase in the number of derivatives that must be taken. More problematically, we found this method to be less stable than the direct magnetic field-evolving method, especially at the boundary.

It appears that the best approach in the context of a spectral method may be to do nothing, and rely on the accuracy of the derivatives, and hence the smallness of the truncation error, to maintain the solenoidal condition to high precision. Let us define a normalised magnetic divergence,
\beq
\lp \divB \rp_{\rm norm} \equiv \frac{\divB}{|\B|/\sqrt{A}},
\eeq
where $A$ is the cell area\footnote{Our discretisation is based on nodes rather than cells, but here it is unimportant which adjacent nodes are chosen to form a fictitious cell.}. Our results are highly divergence-free, as we illustrate with steady-state solutions to three problems. For the Michel rotating monopole (Section~\ref{sec:michel}), the normalised magnetic divergence is $\sim 10^{-14}$ for an $84\times 56$ grid, and $\sim 10^{-18}$ for a grid of $192 \times 128$. The twisted magnetosphere (Section~\ref{sec:twist}) has normalised $\divB$ mostly around $\sim 10^{-12}$, rising to $10^{-9}$ in the region with largest current. Our fiducial aligned rotator solution (Section~\ref{sec:pulsar}) has normalised $\divB$ of $\sim 10^{-6}$ in and near current sheets and $\sim 10^{-10}$ elsewhere; the highest-resolution solution has values roughly ten times lower everywhere.

The higher $\divB$ near current sheets is at least partly due to the presence of the aforementioned Gibbs oscillations, and would be expected to be much smaller in the recoverable high-accuracy solution (see Section~\ref{sec:postprocess}). This Gibbs-generated divergence would also be present if a vector potential had been used. Finally, $\divB$ is higher than normal right at the inner boundary, presumably because of the filtering applied to the Chebyshev derivatives, which become increasingly dependent on the highest frequencies as the boundaries are approached \citep[e.g.][]{Godon:1993p724}.

Most importantly, no unusual or troubling behaviour has been observed to be correlated to an increase of the magnetic field divergence on the grid, the evolution appears to be stable for very long run times, and the measured divergence decreases with increasing resolution.

\subsection{Code infrastructure}

\textsc{phaedra} is written in C for speed and portability. The spectral transforms are performed with the FFTW3 library \citep{Frigo:2005}, which allows transforms of arbitrary size with $O(N \log N)$ complexity. The code is fully MPI-parallel, with a simple automatic domain-decomposition function which does not require the grid dimensions to be a multiple of the number of processors. 

The parallelisation works similarly to the method of \citep{Pelz:1991p647}. The domain, in real space, is slab-decomposed in the radial direction. In the forward transform to spectral space, $f(r,\theta) \rightarrow a_{n l}$, the $\theta$-transforms are first performed using data local to each processor, after which the mixed $f_l(r)$ data undergo a parallel matrix transpose, done with a single \texttt{MPI\_Alltoallv()} call. The $r$-transforms can then be performed, and the coefficients $a_{n l}$, now decomposed in the $l$-direction, used to calculate the coefficients of the desired derivative. The real-space derivative values are found by repeating the above steps in the reverse order.  The filter is applied just before the inverse transform. In the SSV filtering step, performed at the end of each full time step, the coefficients are calculated, filtered, and immediately transformed back to real space. To date the code has been run on between one and 64 processors.

The data output is done collectively, in parallel, using the Parallel HDF5 library. The data are accessed via the XDMF standard, in which an auxiliary XML file, also written by the code, describes the contents of the HDF file containing the data arrays. This format allows the data to be easily opened by the VisIt visualisation software, among others, without the need of a custom plugin.

\section{Test problems}
\label{sec:tests}
\subsection{1D tests}

For the following two test cases we use a one-dimensional simplification of the code, with Chebyshev polynomials as the basis functions, an arcsine coordinate map, and a Cartesian vector basis (which simply requires using the metric $\gamma_{ij} = \delta_{ij}$). The boundary conditions are enforced strongly at $n=0, N-1$; no sponge layers or non-reflecting boundary conditions are used.

\subsubsection{Stationary Alfv\'{e}n wave}

\citet{Komissarov:2004p747} describes an analytical solution for a stationary Alfv\'{e}n wave: $B_x = B_y = E_z = 1$, $E_y = 0$, 
\beq
   B_z(x)=\left\{ 
   \begin{array}{ll} 
    1 & \text{for } x < 0, \\ 
    1 + 0.15 \lb1+\sin \ls 5\pi\lp x-0.1 \rp\rs\rb& \text{for } 0 < x < 0.2, \\ 
    1.3 & \text{for } x > 0.2,  
   \end{array} \right.
\eeq
and $E_x = - B_z$. For ease of comparison with this paper, we also use $N=200$ and a domain $x \in [-1.5,1.5]$. The SSV filter strength is set to $\alpha_{\rm SSV}=0.1$ (even though no such filtering is required for this smooth problem) so that the effect of a discontinuity-capturing level of diffusion can be observed. 

\begin{figure}
\includegraphics[width=84mm]{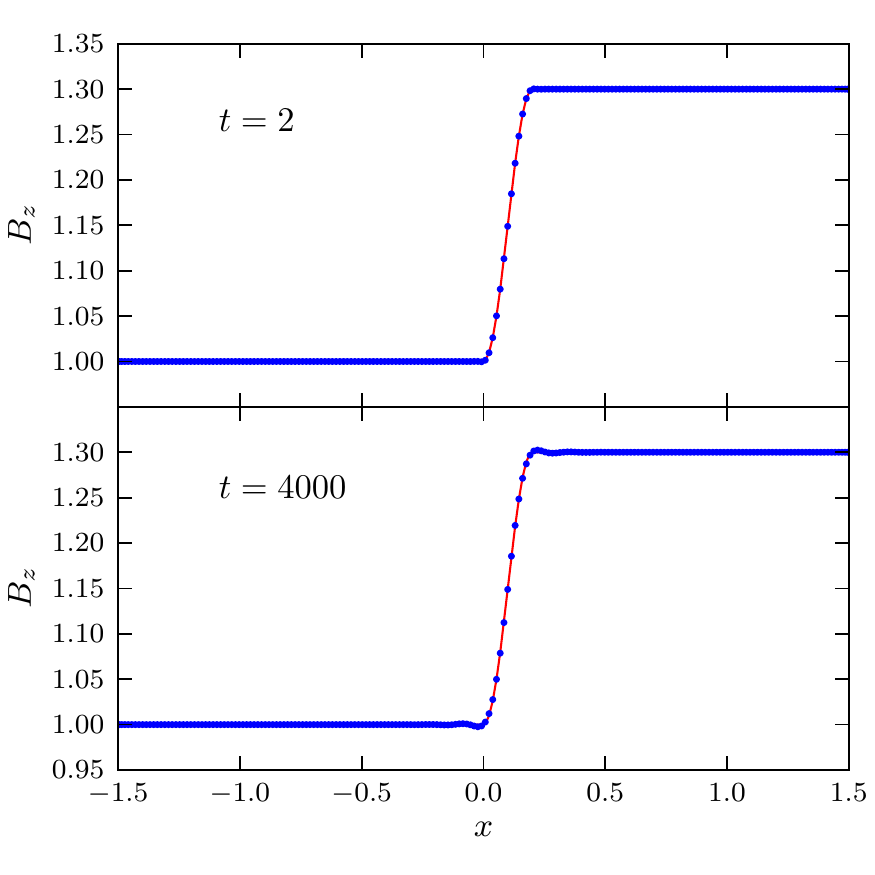}
\caption{Stationary Alfv\'{e}n wave test problem. The line shows the exact solution, the dots show the numerical solution.\label{fig:alfven}}
\end{figure}

Fig.~\ref{fig:alfven} shows the numerical solution for $B_z$ at $t=2$ and $t=4000$. The profile does not broaden noticeably, even after more than a million time steps. The small wiggles, on either side of the jump in the $t=4000$ solution, are imprinted by the action of the SSV filter on a function which is not infinitely smooth ($B_z$ has discontinuous first derivatives at $x=0$ and $x=0.2$).

\subsubsection{Riemann problem}

A Riemann problem which results in a current sheet is described by \citet{Komissarov:2004p747}. The initial conditions are: $\E = \vec{0}$, $B_z = 0$, $B_x = 1$, and 
\beq
B_y(x) = \lb 
  \begin{array}{ll}
   \;\;\; B_0  & \textrm{for } x<0, \\
    -B_0 & \textrm{for } x>0.
  \end{array}
  \right.
\eeq
The current sheet forms spontaneously at $x=0$, and two fast step waves are emitted, one in either direction. The numerical solution for $B_y$ at $t=1$ is shown in Fig.~\ref{fig:riemann}; $\alpha_{\rm SSV} = 0.1$, and $2p_{\rm SSV} = 8$ as usual. The three jump discontinuities are clearly unresolved on the grid, and the fast waves remain sharp. The Gibbs oscillations are confined to the immediate vicinity of a discontinuity, and the affected region shrinks as resolution is increased.

\begin{figure}
\includegraphics[width=84mm]{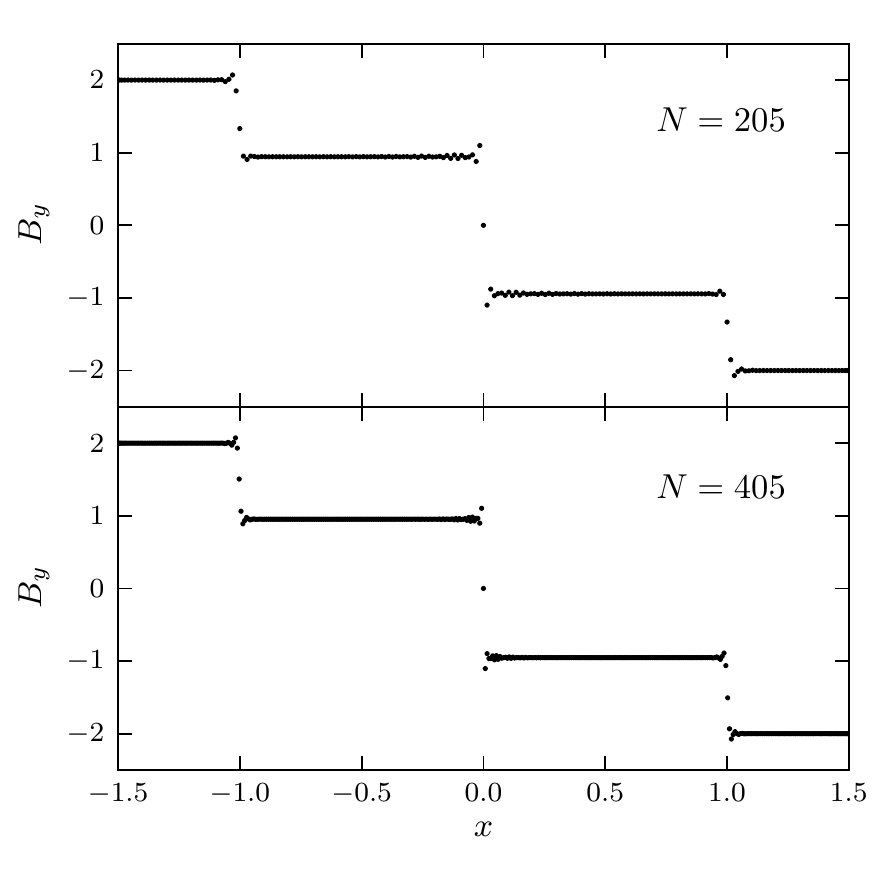}
\caption{Riemann problem at $t=1$, with current sheet at $x=0$, for two grid sizes $N$.\label{fig:riemann}}
\end{figure}

\subsection{2D tests}

In these problems we used the full 2D code, in spherical coordinates. In all cases the star has a radius of unity, $\rin = 1$; its radial light-crossing time-scale is then $\Delta t = 1$. 

\subsubsection{Michel  monopole \& split monopole}
\label{sec:michel}
The exact three-dimensional solution for the field configuration, in force-free electrodynamics, surrounding a rotating magnetic monopole was derived by \citet{Michel:1973}. Given initial conditions $B^r = f_0/r^2$, $B^{\theta} = B^{\phi} = 0$, $\E=0$, and subject to uniform rotation with angular velocity $\Omega$, the steady-state solution has a toroidal magnetic field
\beq
B_{\phi} = f_0 \, \Omega \frac{\sin\theta}{r}.
\eeq

This analytic solution is well suited to a multi-dimensional convergence test. We set the domain to be $1 \leq r \leq 30$ and vary the grid size $N\times L$, holding $N \approx 1.5 L$. No sponge layer is used, since monopole field lines imply radial outgoing waves, for which the characteristic boundary treatment is very effective (Section~\ref{sec:outerboundary}). The order of the aliasing-controlling filter is $2p = 36$ for $L \geq 32$, and slightly lower for smaller grid sizes; no SSV filtering is applied because the solution is smooth. The angular velocity is smoothly increased from zero to $\Omega=0.1$ in twenty radial light-crossing times (i.e. between $t=0$ and $t=20$), and the solution is sampled at $t=100$, on a surface of constant radius at $r=5$. The fractional errors in $B_{\phi}$, defined as
\beq
\textrm{fractional error} = \left|\frac{B_{\phi}^{\rm numerical}}{B_{\phi}^{\rm analytic}} - 1\right|,
\eeq
are plotted in Fig.~\ref{fig:michel-mono}. The errors decrease approximately exponentially with increasing resolution, reaching a level of roughly $10^{-12}$ with a grid size of $84 \times 56$. No effort was made to optimise the domain size, or the radius or time at which the numerical and analytic solutions are compared.  

\begin{figure}
\includegraphics[width=84mm]{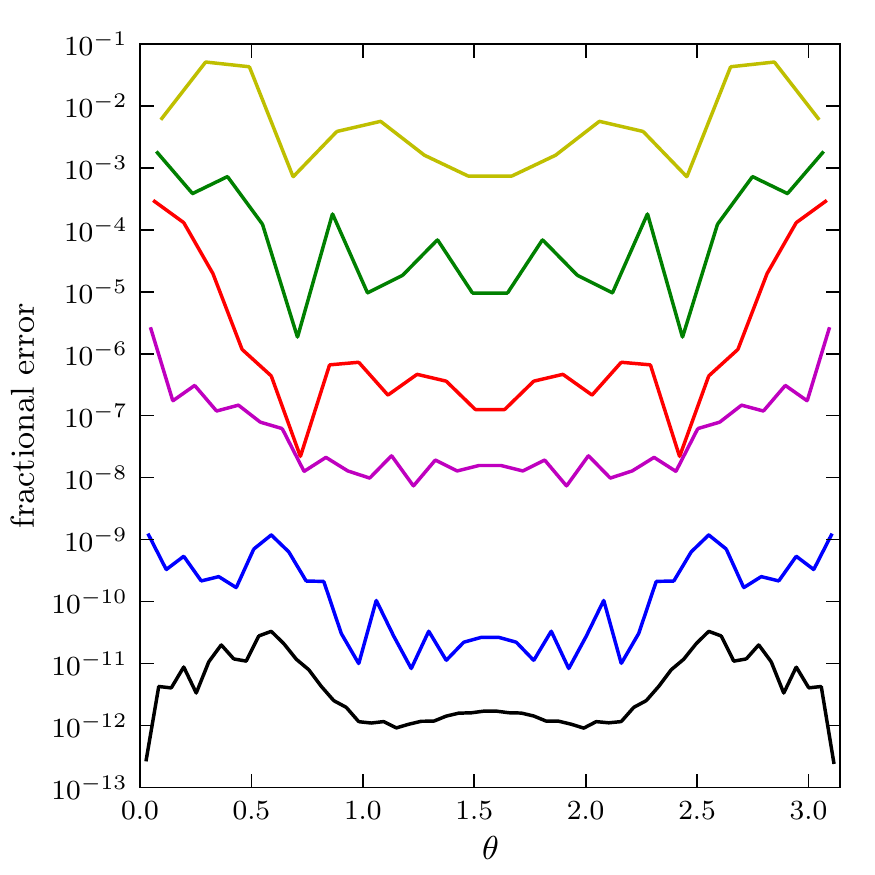}
\caption{Fractional errors in $B_{\phi}$ for a rotating monopole. Grid sizes are, from top, 
$L=16, 20, 24, 32, 40, 56$; $N \approx 1.5 L$. $N$ and $L$ are the number of nodes in the radial and angular directions, respectively. \label{fig:michel-mono}}
\end{figure}

A similar test can be performed for the split monopole, which simply involves reversing the sign of $f_0$, and therefore of $B^r$ and $B^{\phi}$ in the solution, across the equator. The discontinuous magnetic field implies the existence of an equatorial current sheet.  This configuration allows us to test the behaviour of the code in the presence of a realistic discontinuity; we use SSV filtering with $\alpha_{\rm SSV} = 0.05$. Fig.~\ref{fig:michel-splitmono} shows the errors for the split monopole, for a problem otherwise identical to that described previously. 

\begin{figure}
\includegraphics[width=84mm]{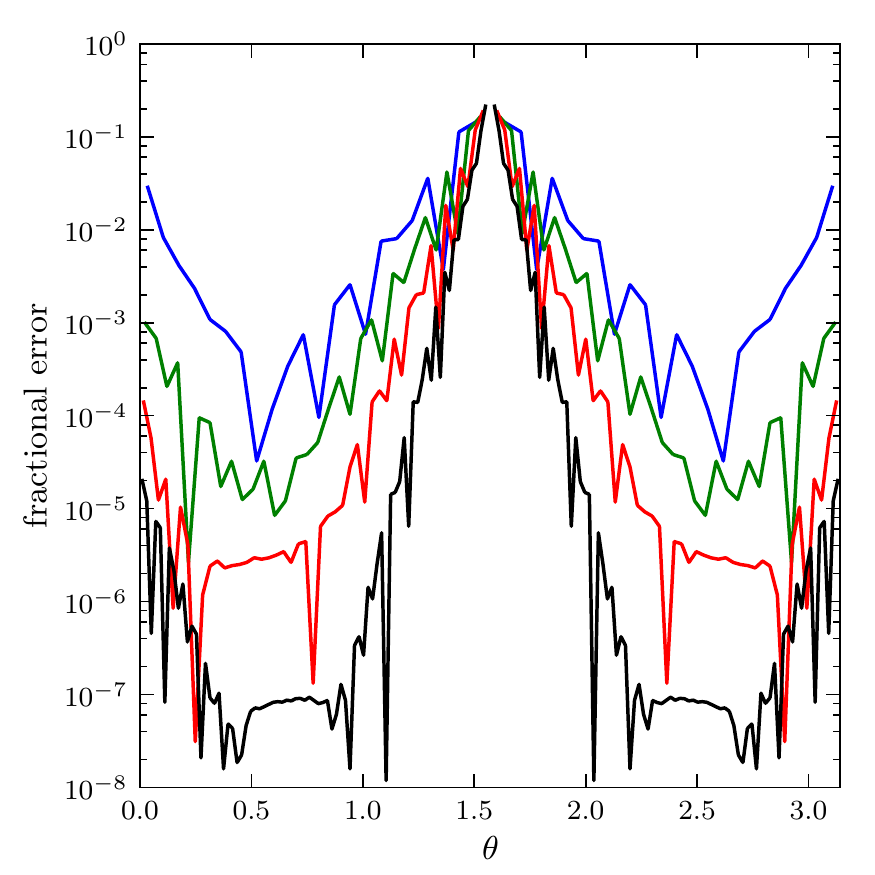}
\caption{Fractional errors in $B_{\phi}$ for a rotating split monopole. Grid sizes are, from top, 
$L=45, 65, 95, 155$; $N \approx 1.5 L$. \label{fig:michel-splitmono}}
\end{figure}

In this case the convergence is not uniform, being faster farther from the discontinuity at the equator. Order unity errors remain immediately beside the current sheet, as described in Section~\ref{sec:filtering}, but the solution converges strongly everywhere else, reaching a fractional error floor of $10^{-7}$ for the $224 \times 155$ grid. The large pointwise errors near the equator are not worrying, because they are due to the superposition of well-behaved, understood, and controlled Gibbs oscillations on top of an otherwise accurate solution, rather than uncontrolled numerical noise or error; see Section~\ref{sec:postprocess} for a discussion.

\subsubsection{Twisted dipole}
\label{sec:twist}

As another 2D test, consider a dipole magnetosphere that is twisted
by a differential rotation of the star's surface.
We assume a latitude-dependent surface motion that is symmetric
about the dipole axis; then the magnetosphere remains axisymmetric.
The twisted field lines become extended in the azimuthal direction
between their footpoints in the northern and southern hemispheres.

We start from a normal dipole configuration at $t=0$ and gradually impart
a twist by moving the surface; at $t=T$ the surface motion stops. If the
magnetosphere is an ideal conductor, it must remain static at $t>T$ --- the
twist will remain `frozen' and persist forever. In the presence of resistivity,
the magnetosphere must untwist with time \citep{Beloborodov:2009}.
Our ideal force-free model has no physical dissipation, and the rate of
untwisting at $t>T$ will measure the numerical dissipation for our scheme.

The speed of the surface motion is much smaller than the speed of light, and so the twist will be implanted
slowly, relative to the relevant wave-crossing time-scales.
In our simulation the stellar rotation is applied only in an annular region of one hemisphere:
\beq
\Omega(\theta) =  \frac{\Omega_{\rm centre}}{1 + \exp \lb -\kappa \ls g \lp \theta - \theta_{\rm c} \rp + \theta_{\rm HW} \rs \rb},
\eeq
where $\theta_{\rm c}$ is the colatitude of the centre of the annulus, $\theta_{\rm HW}$ is its angular half-width, $\kappa$ determines the sharpness of its edges, and
\beq
g =
\begin{cases}
1& \text{if  $\theta < \theta_{\rm c}$},\\
-1& \text{if  $\theta > \theta_{\rm c}$}.
\end{cases}
\eeq
The rotation rate is increased smoothly from zero to $\Omega_{\rm max}$ in time $T/2$,
and returned symmetrically to zero,
\beq
\Omega_{\rm centre} =
\begin{cases}
   \lp \Omega_{\rm max}/2\rp \ls 1 - \cos \lp 2\pi t/T \rp \rs & \text{if $t \leq T$}, \\
   0 & \text{if $t > T$},
\end{cases}
\eeq
twisting the affected region of the star, $\theta_{\rm c} - \theta_{\rm HW} \leq \theta \leq \theta_{\rm c} + \theta_{\rm HW}$, through an angle of $\psi = \Omega_{\rm max} T/2$.

Here we use a grid of $N \times L = 320 \times 255$ on a domain $r \in [1, 40]$, and twist an annular region of the northern hemisphere, given by $\theta_{\rm c} = 0.14\pi$, $\theta_{\rm HW} = 0.04\pi$, and $\kappa = 60$. For this test we want only a small perturbation on top of a dipole field (small total twist amplitude), and for the magnetosphere to move slowly through a sequence of quasi-equilibrium states (requiring small $\Omega_{\rm max}$ and large $T$), and so choose $\Omega_{\rm max} = 10^{-3}$ and $T=200$.
This implants a twist of $\psi = 0.1$~radians by time $t=200$, after which the stellar surface is at rest.

To highlight the importance of the parallel current (see Sec.~\ref{sec:current}) we perform three simulations, all using the usual eighth-order SSV filtering with $\alpha_{\rm SSV} = 0.05$. In run (a) we use the full force-free current in the equations of motion, $\J = \J_{\parallel} + \J_{\rm drift}$ from equation~(\ref{eq:J2}); in (b) we keep only the drift current, $\J = \J_{\rm drift}$, and remove accumulated $\E_{\parallel}$ at every Runge-Kutta substep; in (c) we use only the drift current, and remove parallel electric field at the end of every full time step. Run (a) therefore uses our standard numerical scheme; for this simulation the profile of $B_{\phi}$ along the equator at $t=200$ is shown in Fig.~\ref{fig:twistB3}.

\begin{figure}
\includegraphics[width=84mm]{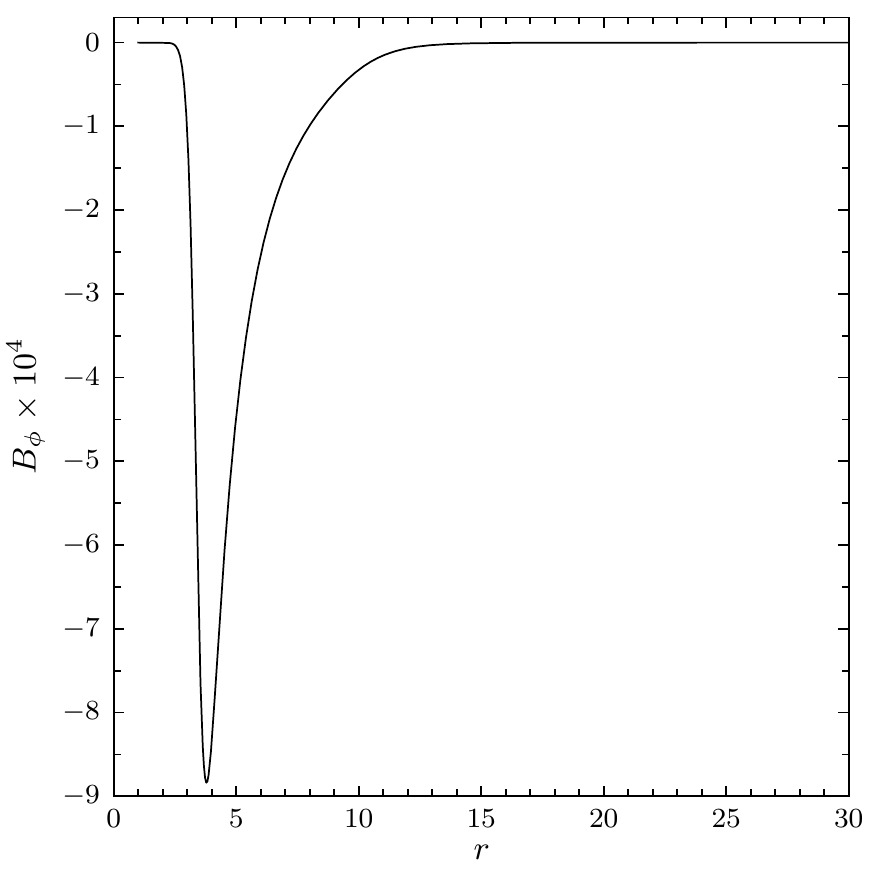}
\caption{ $B_{\phi}$ versus $r$ at $\theta = \pi/2$, at $t=200$, once the twist profile has been implanted; simulation (a) (see text).\label{fig:twistB3}}
\end{figure}

\begin{figure*}
\includegraphics[width=\textwidth]{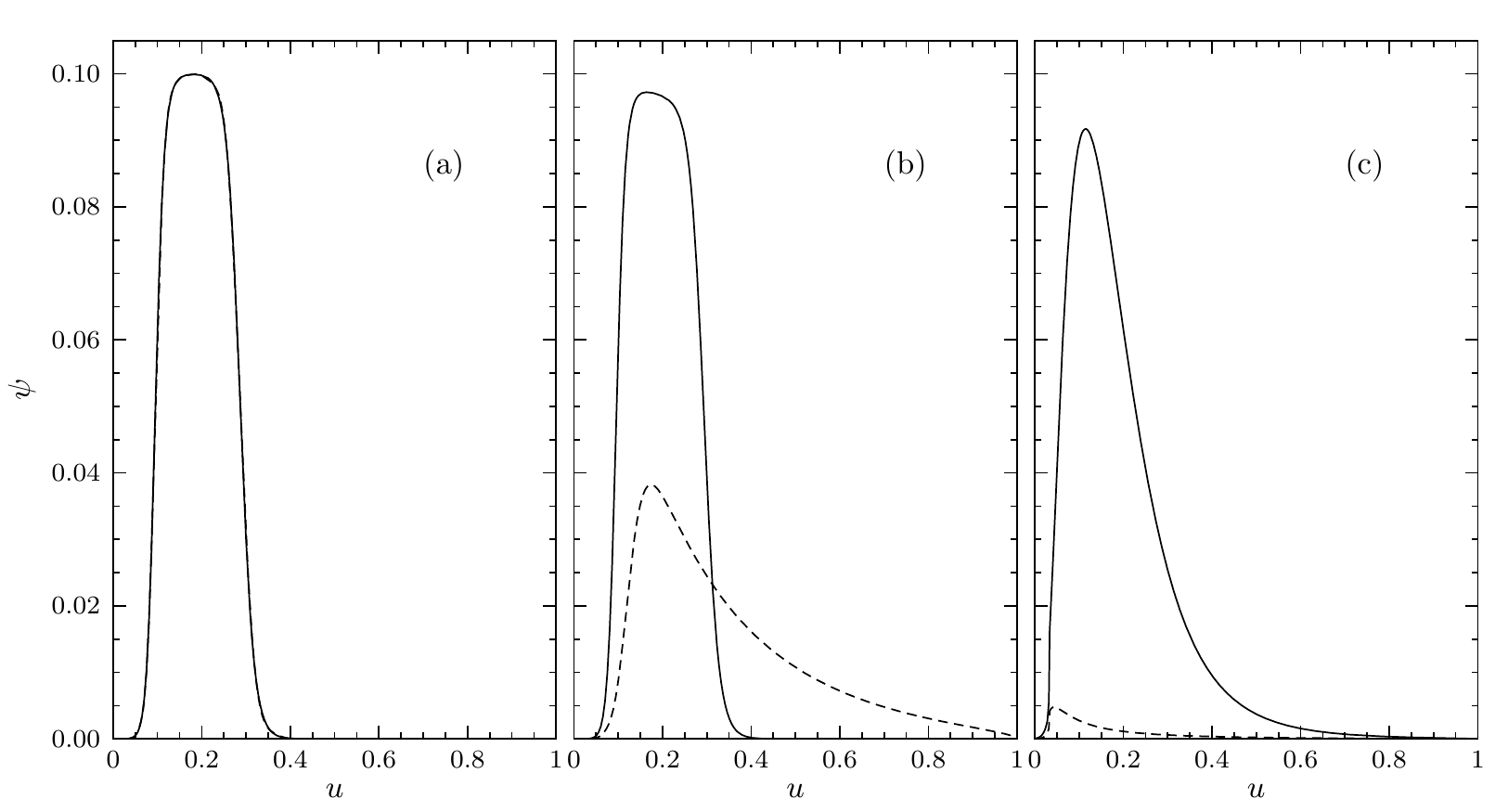}
\caption{Twist versus flux function at $t=200$ (solid line) and $t=4000$ (dashed line), using (a) full current density, (b) drift current only, enforcing the $\E\cdot\B=0$ constraint at each Runge-Kutta substep, (c) drift current only, enforcing the constraint at the end of a full time step. \label{fig:twistcompare}}
\end{figure*}

\begin{figure}
\includegraphics[width=84mm]{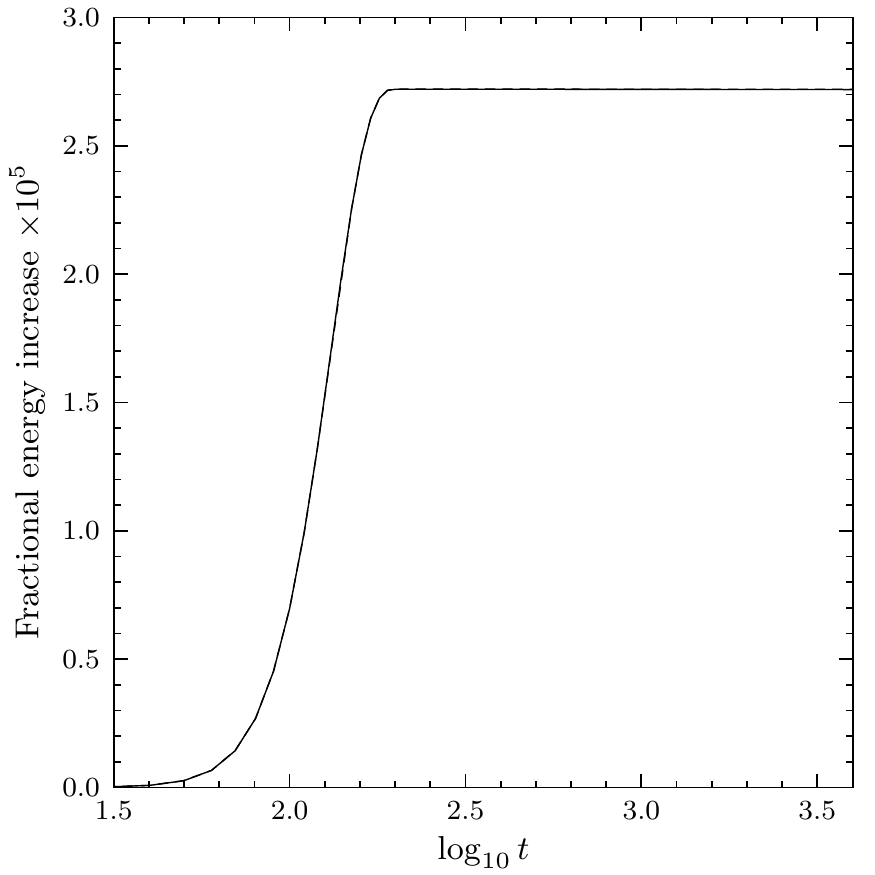}
\caption{ Energy injected into magnetosphere by twisting, in units of initial field energy: $W_1 - W_0$ (solid line), $W_2$ (dashed line). The lines lie on top of one another. \label{fig:energy}}
\end{figure}

The total twist along any field line can be calculated by numerically integrating its path in space: the twist is $\psi = |\phi_{\rm end} - \phi_{\rm start}|$. We label the field lines with the fractional flux function, $u$. It measures the fraction of the star's magnetic flux passing through a circular contour, centred on the magnetic axis, which goes through the field line's footpoint. This function is $u = \sin^2 \theta_1$ for a dipole field, where $\theta_1$ is the co-latitude of the northern footpoint, a relationship which is unchanged by any axisymmetric motion of the stellar surface.

Fig.~\ref{fig:twistcompare} shows the measured twists at two times, $t=200$ (solid lines) and $t=4000$ (dashed lines), for each of the three runs. In run (a) the curves lie on top of one another; in fact, sharp twist profiles are preserved, at this resolution, even for tens of thousands of light-crossing times. In run (b) the profile at $t=200$ is fairly good, but it has diffused significantly by $t=4000$, while in run (c) the profile is already inaccurate by the earlier time and has almost completely disappeared by the later. These three cases demonstrate how important the parallel current is for self-consistently maintaining $\E\cdot\B=0$ throughout each time step; omitting it implies making errors which demand the removal of electric field from the system, introducing an artificial source of dissipation.

We focus now on run (a). The permanence of the implanted twist seen in Fig.~\ref{fig:twistcompare} can also be interpreted in terms of energy conservation. Define $W_1$ as the total electromagnetic energy of the twisted configuration, instantaneously measured on the grid,
\beq
W_1 = \frac{1}{2}\int_V \lp B^2 + E^2 \rp {\rm d}V,
\eeq
where $V$ is the three-dimensional volume of the computational domain. Let $W_0$ be the energy for the initial pure dipole field. Define $W_2$ as the energy calculated by numerically integrating the net energy flux into the grid over time,
\beq
W_2 = \int_0^t \ls L_{\rm in}(t') - L_{\rm out}(t') \rs {\rm d} t',
\eeq
where
$L = 2\pi r^2 \int_0^\pi (\E \times \B)_r \sin\theta\, \mathrm{d}\theta$, and the subscripts denote the luminosity measured at the inner and outer boundary surfaces respectively. The part of the grid containing the absorbing layer is not included in calculating $W_1$ or $W_2$. If there is no dissipation, we expect $W_1 - W_0 = W_2$.

The measured twist energy $W_1 - W_0$ and the imparted energy $W_2$ are shown in Fig.~\ref{fig:energy} as a function of time. The two curves lie on top of one another to within their widths. The energy on the grid is stable once the twist-up phase is complete. Between $t=200$ and $t=4000$ the twist energy decreases by a factor of  $4\times 10^{-4}$. At $t=4000$, the fractional difference between accumulated twist energy and integrated net Poynting flux is $\ls \lp W_1 - W_0 \rp - W_2 \rs / \lp W_1 - W_0 \rp = -3.9 \times 10^{-4}$.

\section{The aligned rotator}
\label{sec:pulsar}

The objects which we observe as radio pulsars are generally accepted to be rotating magnetised neutron stars, with magnetic fields of $10^{12}$ G or more. If the star's basic magnetic field is modelled as a dipole, the electric field induced by its rotation, equation~(\ref{eq:perfconduct}), has a large radial component, which is able to rip charged particles from the stellar surface \citep{Goldreich:1969p4440}. These particles, and e$^\pm$ pairs created in the magnetosphere, surround the star with force-free plasma.

To simplify the problem, one can consider the axisymmetric configuration with the magnetic moment parallel to the rotation axis: the aligned rotator. A steady-state solution was found by \citet{Contopoulos:1999p3926}, which included a region of closed field lines extending to the light cylinder (defined as the cylindrical radius $\rlc$ at which the co-rotation speed is the speed of light), and open, asymptotically monopolar, field lines extending to infinity. A current sheet is present at the equator beyond the light cylinder, which splits at the `Y-point' to follow the last closed flux surface. It was later found that equilibrium solutions exist with the Y-point at any distance from the rotation axis, within the light cylinder \citep{2004MNRAS.349..213G, 2005A&A...442..579C, Timokhin:2006p4512}. Absent a unique solution, one turns to time-dependent studies \citep{Spitkovsky:2006p752, McKinney:2006p974, Komissarov:2006p978, Kalapotharakos:2009p3903, Yu:2011p4007}, which have all found that the Y-point moves toward the light cylinder, where it subsequently remains.

\subsection{Numerical setup}

In our simulations, spatial and temporal scales are fixed by setting $r_* = 1$, where $r_*$ is the stellar radius; $c=1$ as usual. The star is smoothly spun up from rest, over a few light-crossing times, to the rotation frequency $\Omega$, implying a light cylinder at $\rlc = 1/\Omega$. We have investigated cases with $\rlc = 5$, 10, 20, 30, and 40;  there were only minor differences between the solutions, and here we concentrate on those with $\rlc = 20$. In these we set $\rout$ = 70, with an absorbing layer beginning at $r$ = 50. A radial arcsine-plus-algebraic coordinate mapping is used, with parameters $\epsilon = 10^{-11}$ and $Q = 0.7$; the grid is equally spaced in the meridional direction. The current sheets are captured with super spectral viscosity (SSV): $2 p_{\rm SSV} = 8$, $\alpha_{\rm SSV} = 0.1$. 

We have performed simulations with a range of grid sizes, from $N \times L$ = $81 \times 49$ to $768 \times 507$. In order to demonstrate that a very fine grid is not needed for accurate results, we will concentrate on our run with a grid of $384 \times 255$---all the following results are from this simulation unless explicitly noted otherwise. The behaviour is similar for all resolutions. 

The initial magnetic field is set to a dipole with unit magnetic moment: $B_r = 2 \cos\theta/r^3$, $B_{\theta} = \sin\theta/r^3$, $B_{\phi} = 0$. Rotation is introduced by applying an electric field, equation~(\ref{eq:perfconduct}), bringing the star to its final angular velocity by $t=10$.

\subsection{Evolution to steady state}

During the spin-up phase, Alfv\'{e}n waves are injected along field lines into the magnetosphere, filling it with charges and currents. The magnetic energy increases and the poloidal field lines inflate, appearing to be pulled outward along the equator, where waves from opposite hemispheres meet. By $t=40$ a clear distinction can be seen between those lines which are too close to the star to be strongly inflated, and those which are on the path to opening. Waves can be seen to propagate backwards and forwards on the former, while the latter are expanding at nearly the speed of light. Poloidal field line projections are drawn in Fig.~\ref{fig:flines}.

\begin{figure*}
\includegraphics[width=170mm]{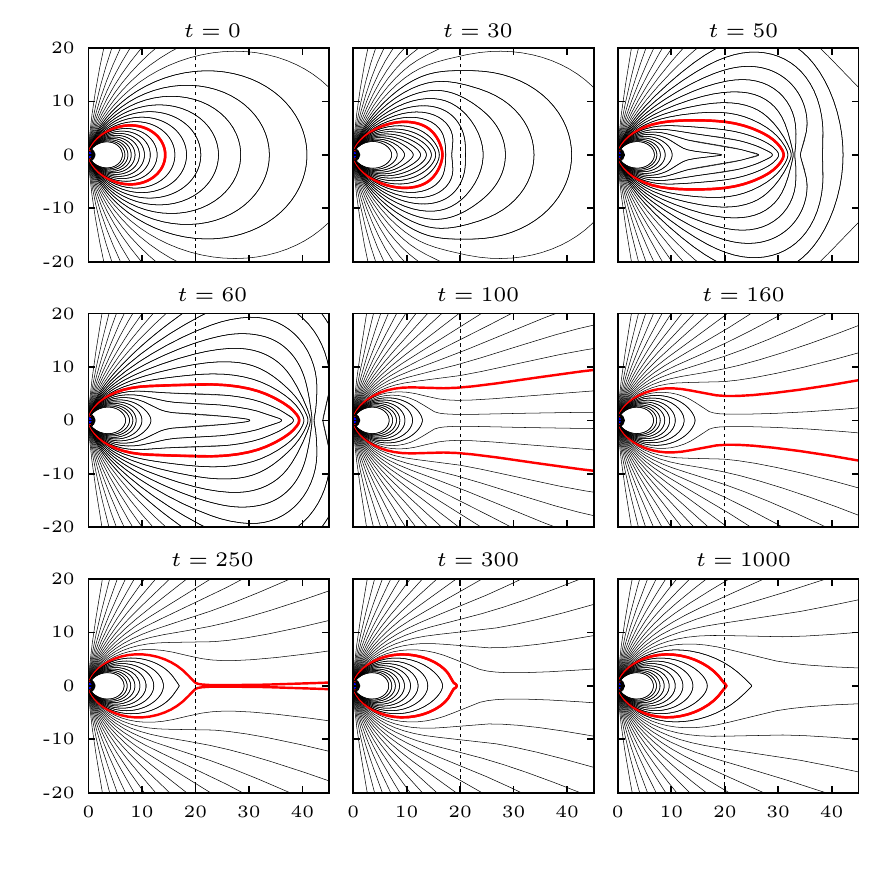}
\caption{Poloidal field lines for the aligned rotator, from non-rotating at $t=0$ to the equilibrium solution at $t=1000$. The thicker, red field line closes at the Y-point in the equilibrium solution; the dashed line at $R=20$ indicates the light cylinder. There are 25 lines drawn from each pole, evenly spaced in colatitude between $0.01\pi$ and $0.13\pi$. \label{fig:flines}}
\end{figure*}

Let us define the closed zone as that region with oscillating fields and currents, and the Y-point as the point at which the currents around the closed zone meet at the equator. All the fields are smooth before $t \approx 35$, at which time the radial and azimuthal currents collapse to an unresolved equatorial current sheet. As the configuration evolves, all components of the current around the closed zone, and in the equatorial current sheet, change direction, as one can see in Fig.~\ref{fig:ypoint}. The Y-point is at $r = 0.5\rlc$ before the reconfiguration, and at $r=0.6\rlc$ afterwards. By $t=90$ Alfv\'{e}n waves have again filled the closed zone up to the Y-point, and the quasi-steady march of the Y-point to the light cylinder begins.

\begin{figure*}
\includegraphics[width=160mm]{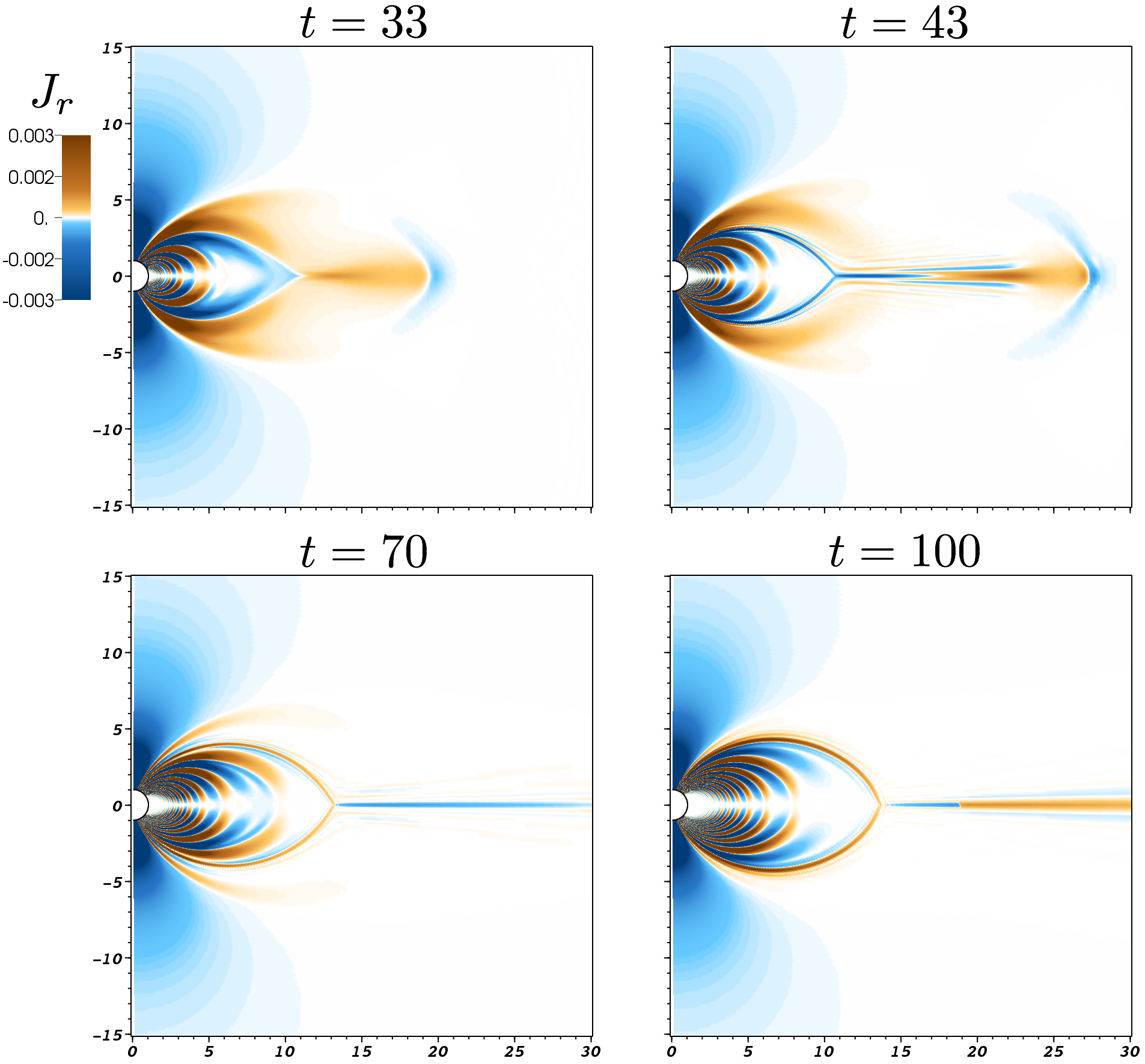}
\caption{Formation of the Y-point. The colour shows radial current density; gold is positive, blue is negative. Charge flows into, and accumulates at,  the Y-point, until the Y-point reaches the light cylinder. The light cylinder is at $R=20$. \label{fig:ypoint}}
\end{figure*}

The equatorial current sheet is unresolved on the grid, having discontinuities in both $B_r$ and $B_{\phi}$; Fig.~\ref{fig:postproc}(a) shows $B_{\phi}$ against $\theta$ at $r=22.25$. The total magnetic field drops close to zero directly on the equator, and so the magnitudes of any electric fields must be reduced so that the second force-free condition, equation~(\ref{eq:b2e2}), is not violated (see Section~\ref{sec:current}). When choosing the grid, it is advantageous to have a line of nodes on the equator, otherwise the current sheet must choose which of the two equidistant nodes to collapse onto---since neither choice obeys the symmetry of the problem, the sheet will periodically move from one line to the other.

At $t \sim 100$, more open flux has been created than can be supported by the energy being pumped into the magnetosphere by the star's rotation. The  Y-point moves very slowly towards the light cylinder, as some of the open field lines reconnect in the current sheet. The Alfv\'{e}n  waves in the closed zone are eventually damped by numerical dissipation\footnote{The waves are sheared by field line curvature, becoming longer and thinner. Numerical dissipation becomes significant when they approach the grid scale, and are attenuated by the filters.}; Fig.~\ref{fig:rescompare} shows the toroidal magnetic field for low ($256 \times 155$) and high ($768\times507$) resolution runs at $t=250$. The oscillations survive longer with higher resolution, but the Y-point moves only slightly slower; here, it is at $r=16.8$, versus 17.5 for the low-resolution run.

\begin{figure*}
\includegraphics[width=160mm]{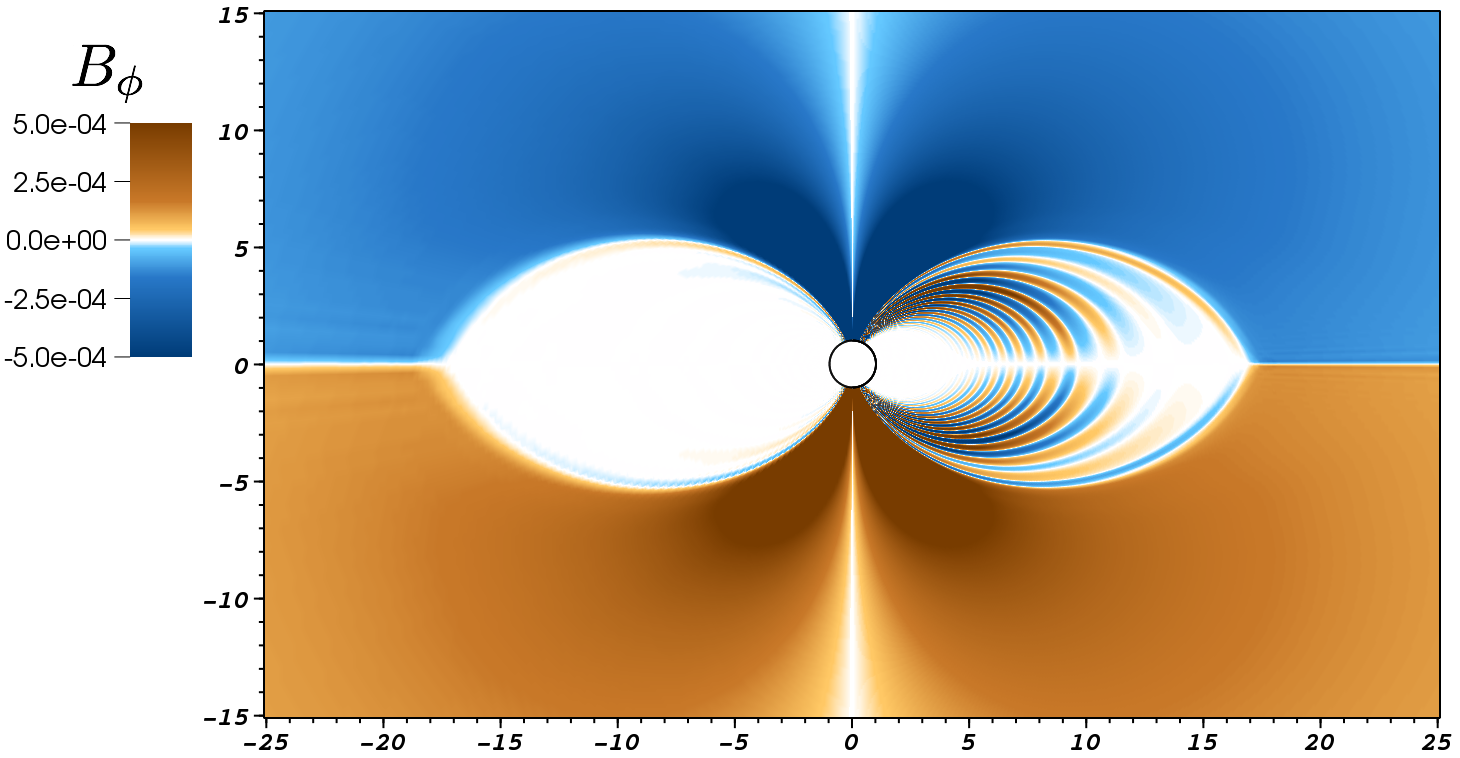}
\caption{Comparing low resolution ($256 \times 155$), left, and high resolution ($768 \times 507$), right, simulations at $t=250$. The colour is toroidal magnetic field. The Y-point has not yet reached the light cylinder at $R=20$. \label{fig:rescompare}}
\end{figure*}

The Y-point reaches the light cylinder at $t \sim 500$, but open flux continues to reconnect slowly until $t \sim 700$, about 5.5 rotation periods. After this time the solution is stationary. The time taken to reach equilibrium is practically independent of grid size. The equilibrium solution has some field lines closed outside the light cylinder due to the effective resistivity of the current sheet; for example see the last panel of Fig.~\ref{fig:flines} or Fig.~\ref{fig:equil-flines} below. Note however that the zero-toroidal-field closed zone is strictly within the light cylinder.

\subsection{Steady state}

We now concentrate on the obtained equilibrium solution. Unlike numerical solutions in previous works, ours does not exhibit plasmoid emission by the Y-point once it has approached the light cylinder. The Y-point and current sheet are steady for long times; we ran a $256\times 155$ simulation until $t=10,000$, or 80 rotational periods, without seeing any indications of Y-point instability. We found plasmoid emission from the Y-point in only two circumstances. Firstly, if the level of filtering was too low; no plasmoids were seen if the filtering was strong enough to prevent the Gibbs oscillations on either side of the current sheets from growing over time. Secondly, if the radial grid spacing was too large near the light cylinder; we found the Y-point was stable if $\Delta r \lesssim 0.75\, r \Delta\theta$. This is probably due to the action of the spectral filters, which make the current sheet mildly resistive, slightly diffusing the Y-point.

The current sheet resistivity largely comes from filtering $B^{\theta}$ in the meridional direction, across the sheet. As the sheet forms, $B^{\theta}(\theta)$ goes to zero at the equator, with a cusp-like profile on each side, implying significant high-wavenumber content. The filters damp these high wavenumbers, causing the smoothed $B^{\theta}$ to be non-zero on the equator, which closes field lines. Eventually an equilibrium is reached between filtering and the electromagnetic forces trying to compress the current sheet. We verified this picture by using unfiltered values of $B^{\theta}$ whenever the $B^2 - E^2 > 0$ condition is violated; this resulted in near-zero magnetic field in the current sheet, and a Y-point that moved outwards much more slowly. However, without filtering the evolution eventually became unstable.

To investigate the dissipation in the current sheet, we performed a simulation using a $576 \times 255$ grid, with an outer boundary at $r=1000$, an absorbing layer beginning at $r=700$, and a more severe coordinate mapping; again, $\rlc = 20$. In Fig.~\ref{fig:poynt-v-r} we plot integrated Poynting flux through concentric spheres, as a function of radius. The outgoing flux is constant within the light cylinder, with maximum fractional variation of about $3\times10^{-4}$ near the star. The flux inside the light cylinder agrees with the value found in previous works, $\mu^2 \Omega^4/c^3$, where $\mu$ is the star's magnetic moment, to fractional accuracy of $6\times10^{-3}$ for our fiducial simulation and $2\times10^{-3}$ for our high-resolution one.

Outside the light cylinder, some of the outgoing flux is lost in the current sheet due to its effective resistivity due to the filters; this deficit decreases with increased resolution. The energy loss is relatively large, because the resistivity is confined to the current sheet, which is kept sharp by the fully ideal surrounding magnetosphere. Solutions with global resistivity dissipate a smaller fraction of their luminosity in the current sheet \citep{Kalapotharakos:2011p4587}. As shown in Fig.~\ref{fig:farlines}, the open field lines are asymptotically radial.

\begin{figure}
\includegraphics[width=84mm]{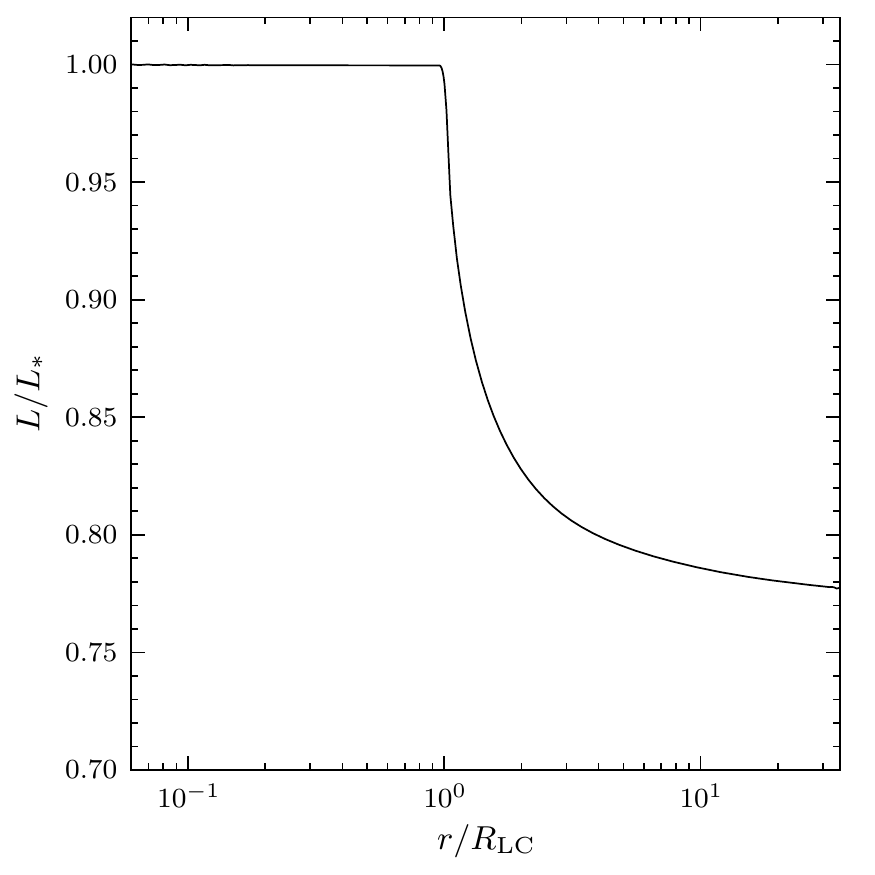}
\caption{Luminosity (integrated Poynting flux) through a sphere of radius $r$, as a function of $r$. $L_*$ is the luminosity measured at the stellar surface; it equals $\mu^2 \Omega^4/c^3$ (to a fractional accuracy of $6\times 10^{-3}$). \label{fig:poynt-v-r}}
\end{figure}

\begin{figure}
\includegraphics[width=84mm]{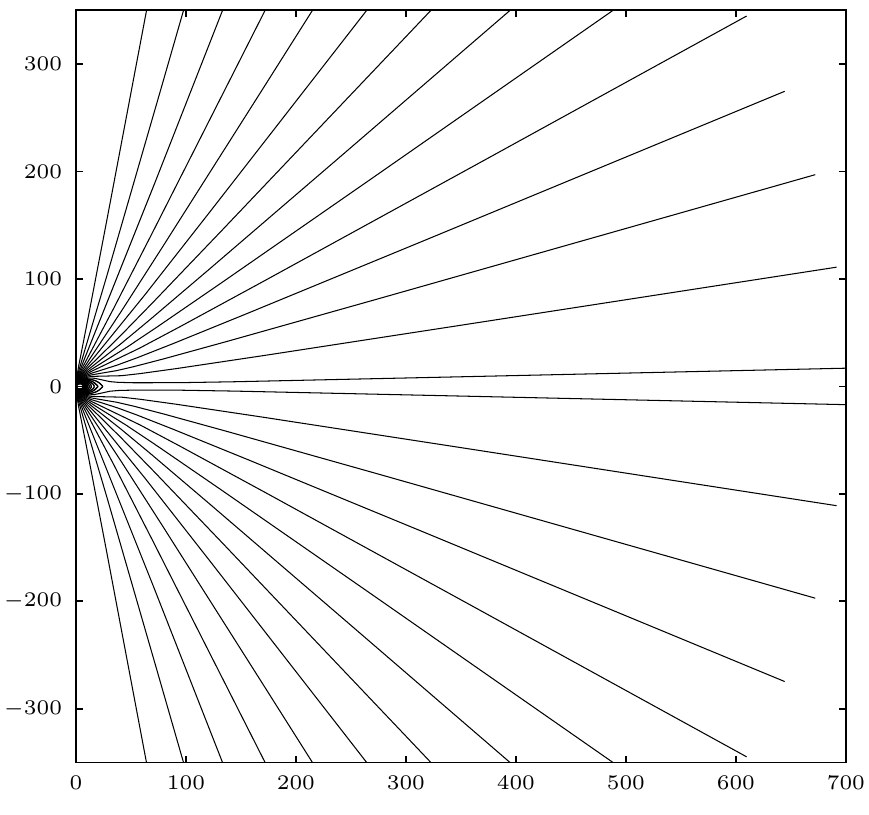}
\caption{Poloidal field lines, out to $r =  35 \rlc$. The field lines have the same footpoints as in Fig.~\ref{fig:flines}. \label{fig:farlines}}
\end{figure}

\begin{figure}
\includegraphics[width=84mm]{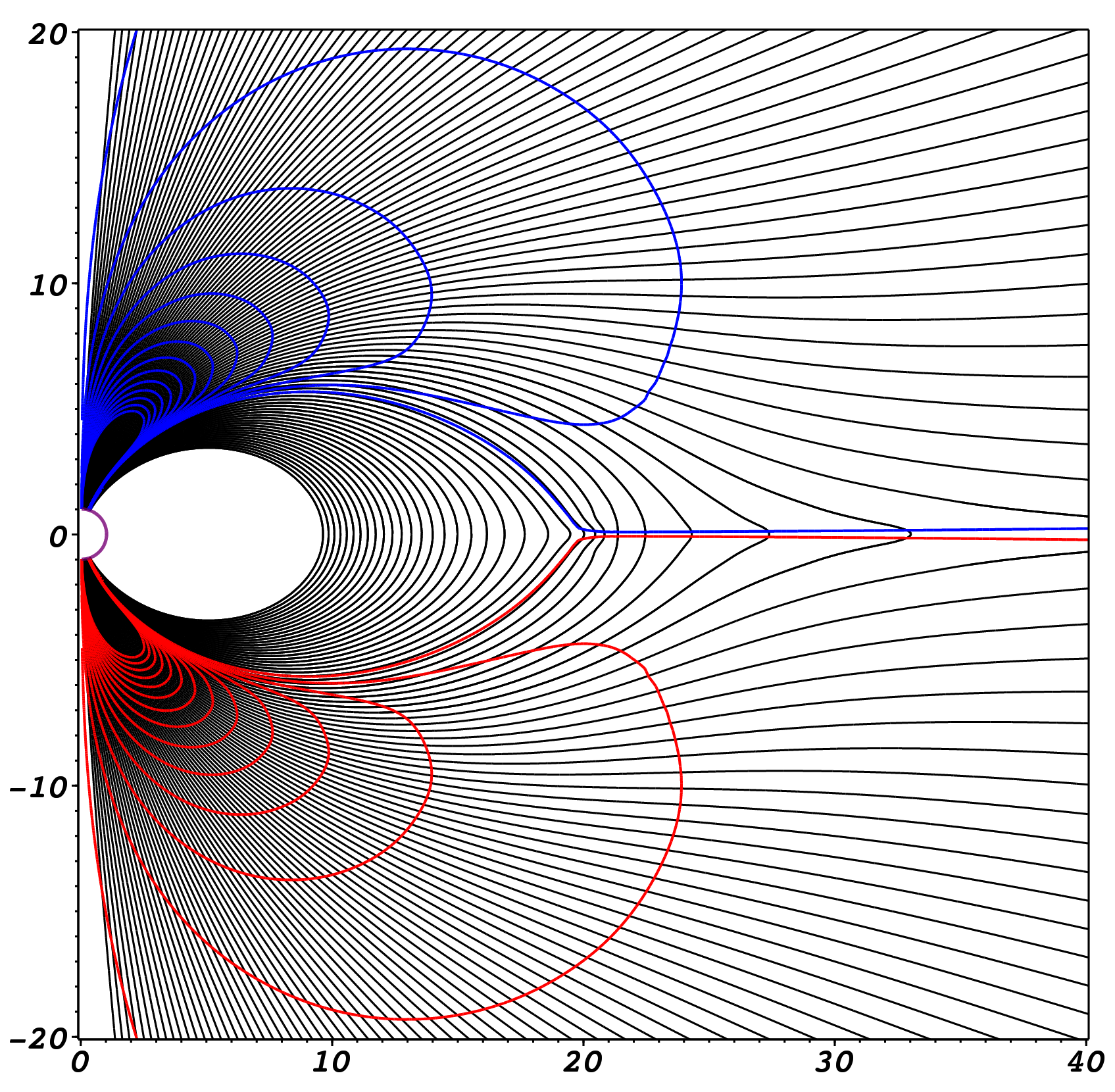}
\caption{Poloidal field lines and contours of constant $B_{\phi}$. There are 100 lines drawn from each pole, equally spaced in colatitude between $\theta = 0.005\pi$ and $0.11\pi$.  Contours, 14 in each hemisphere, are drawn in blue for negative values and red for positive values. They show $|B_{\phi}|$ equally spaced between $3\times10^{-5}$ and $10^{-3}$. \label{fig:equil-flines}}
\end{figure}

The magnetic field lines closing inside the Y-point form the closed zone,
with no poloidal currents and zero toroidal magnetic field. The density of
field lines is higher around the boundary of the closed zone. This can be seen in Fig.~\ref{fig:equil-flines}, where we plot field lines and contours of toroidal magnetic field. The closed field lines outside the light cylinder form cusps at the equatorial current sheet (we do not consider these lines to be part of the `closed zone' because particles are not trapped on them, being able to flow out through the current sheet). The total magnetic flux through the light cylinder, in one hemisphere, is $1.39\, \pm\, 0.01\; \mu\Omega/c$. This is larger than the value, 1.23, obtained by \citet{2005A&A...442..579C} and \citet{Timokhin:2006p4512} for ideal steady-state force-free models, because some of the closed flux in our solution has diffused through the light cylinder due to finite resistivity.

The magnetosphere has current leaving the polar cap and, outside the light cylinder, returning to the star mostly in the current sheet. At the Y-point the current sheet splits in two, and follows the boundary of the closed zone. In Fig.~\ref{fig:polarcap} we plot the poloidal current density measured on the star, normalised to the Goldreich-Julian current density (the speed of light times the equilibrium charge density); a similar plot was obtained for the steady-state solution by \citet{Timokhin:2006p4512}. To aid comparison with Fig.~5 of that paper, we have scaled the axis to the width of the polar cap. The current sheet is seen just inside $\theta/\theta_{\rm pc} = 1$.

The three components of the current density, and the charge density, of the equilibrium solution are shown to a common scale in Fig.~\ref{fig:currents}. Some Gibbs oscillations are apparent near the Y-point and beside the outer current sheet, but they are controlled by the filtering and do not appear to cause any problems. The closed zone has no poloidal current, but there is toroidal current from the corotating charge density.

In a steady, ideal, force-free magnetosphere the electric field
has a potential that is constant along the magnetic field lines.
Our solution is everywhere close to this behaviour (except in the
equatorial current sheet where numerical resistivity is significant).
The equipotentials follow the shape of magnetic field lines, and
are smooth and accurate. The charge density corresponds to the
Laplacian of the potential, and the second derivatives have more
numerical noise. Nevertheless, the obtained charge density
reproduces all the expected features, including steep gradients
near the current sheet and the singular behaviour at the Y-point.
The current sheet is positively charged outside the Y-point and
negatively charged around the closed zone, in agreement with the
previously obtained steady-state solutions \citep[see][]{Timokhin:2006p4512}. 
The negatively-charged current sheet around the closed zone appears to be resolved, and the thickness of the negatively-charged region decreases slowly with increased resolution. This suggests a thickening of the current sheet due to finite resistivity, as argued by \citet{Gruzinov:2011p3139}. The thickening must occur due to resistivity near and outside the Y-point, as
dissipation inside the light cylinder is negligible (see Fig.~\ref{fig:poynt-v-r}).

\subsection{ Viability of the force-free model }

The force-free model relies on the availability of charges that
sustain the required charge density and electric currents.
Charges can be pulled out from the star or supplied by e$^{\pm}$ pair creation.
Both mechanisms require a longitudinal voltage, i.e. $\E\cdot\B \neq 0$;
pair creation, in particular, requires a significant voltage.
In most observed pulsars, this voltage is not so large
as to make the force-free approximation unreasonable.
A real danger for the force-free model appears if the required
charge density or current cannot be created.

The electron-positron discharge operates at the polar cap if
$\alpha^{-1} < 1$, where $\alpha \equiv J/c\rho_e$ \citep{2008ApJ...683L..41B}.
This condition is satisfied in the zone of return current ($J_r > 0$),
near the edge of the polar cap where $\alpha < 0$ (Fig.~\ref{fig:polarcap}).
Pairs are created with a high multiplicity and outflow along
the magnetic field lines, screening $\E\cdot\B$. The presence of dense
pair plasma makes the force-free approximation safe in the
return-current zone (except in the current sheet). The boundary
of this zone ($J_r=0$) is shown by the blue curve in Fig.~\ref{fig:chargecontour}.

\begin{figure*}
\includegraphics[width=160mm]{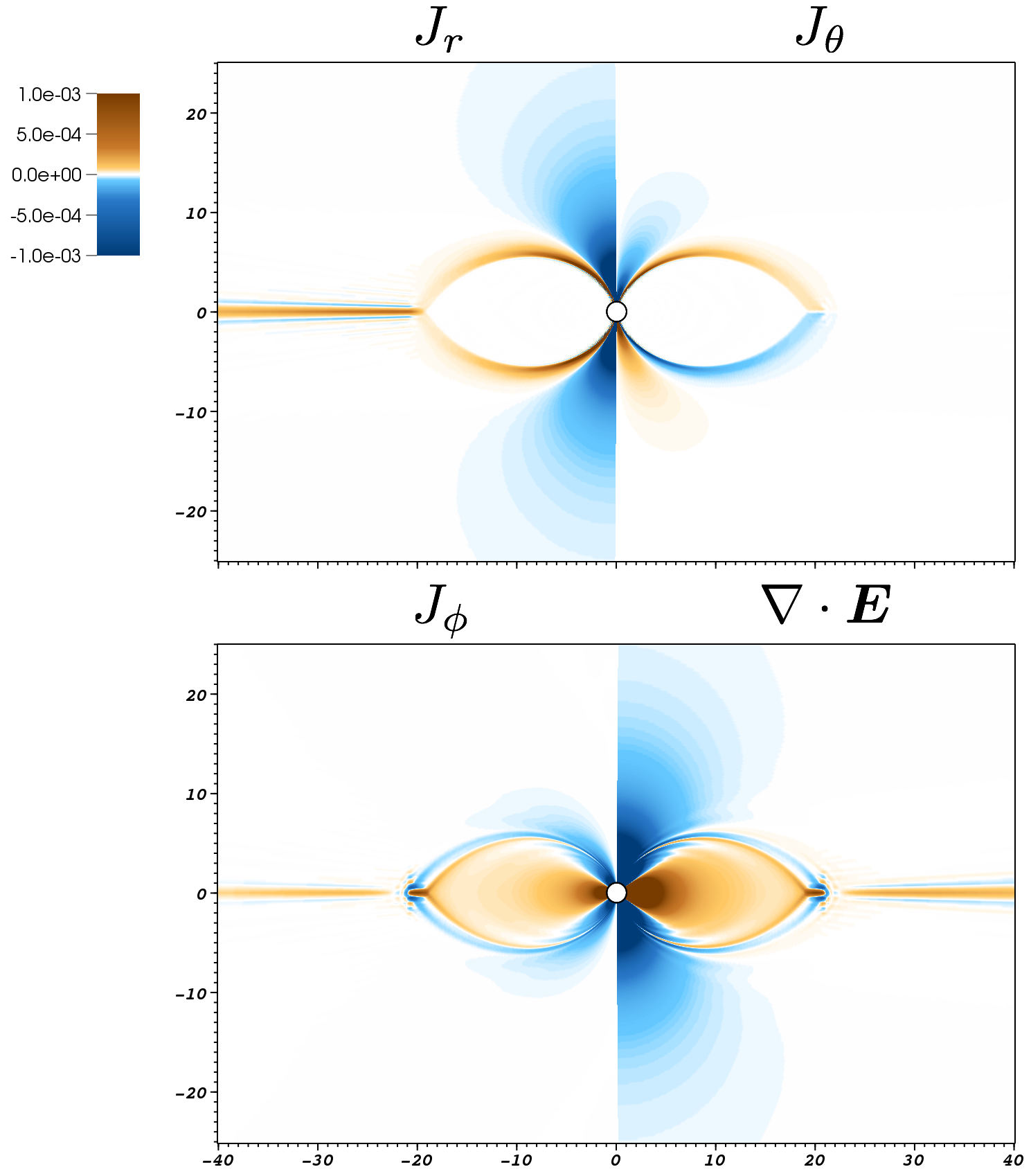}
\caption{Steady-state current, $\J \equiv \curlB$, and electric charge distributions of the equilibrium solution. \label{fig:currents}}
\end{figure*}

The discharge does not occur in the central parts of the polar cap
where $J_r < 0$ and $0 < \alpha < 1$. Instead, the required $\rho_e$ and $J$
are supplied by the charge-separated flow pulled out from the star 
(Beloborodov 2008; Chen \& Beloborodov, in preparation).
The force-free approximation remains accurate
along the field lines extending from this region as long as
$0 < \alpha < 1$ remains satisfied. We find that $\alpha < 1$ everywhere
in the zone $J_r < 0$, inside and outside of the light cylinder.
However, $\alpha > 0$ is {\it not} satisfied. There is a small
region in the zone $J_r > 0$ outside the light cylinder where
$\alpha$ becomes negative (because $\rho_e$ changes sign, see Fig.~\ref{fig:chargecontour}).
The charge-separated outflow passing through this region fails to
supply the charge density of the required sign, and a large $\E\cdot\B$
must develop. $\E\cdot\B$ may be limited by locally initiated pair creation
in young, fast-rotating pulsars; however, in most pulsars pair
creation is inefficient so far from the neutron star.

We conclude that, for the aligned rotator, the force-free
approximation is expected to fail in the region where $J_r < 0$ and
$\rho_e > 0$. This region is, however, small, and this problem may not
impact the obtained global solution.
We note that a similar, but larger, region is seen in Fig.~4 of
\citet{Timokhin:2006p4512} and in Fig.~3 of \citet{Contopoulos:1999p3926}.
The difference between their and our solutions is due to the fact
that their model is strictly ideal everywhere, while our model
is (nearly) ideal only outside the equatorial current sheet.
The dissipation in the current sheet affects the magnetosphere
as discussed above and shrinks the region that is dangerous for the
force-free model outside the current sheet.

We note also that the polar-cap accelerator in the zone of return current $J_r > 0$
can supply some e$^\pm$ pairs to the neighbouring field lines with $J_r < 0$
(near the blue curve in Fig.~\ref{fig:chargecontour}).  Pair creation is not exactly local to the
acceleration region because it involves an intermediate step--the emission
of a high-energy photon which must propagate a finite distance {\it across} the
field lines before converting to e$^\pm$. Pairs created on, and outflowing along,
the field lines slightly outside the return-current zone may supply enough positive
charges to the problematic small region ($J_r < 0$, $\rho_e > 0$) and validate the
force-free condition there.

\begin{figure}
\includegraphics[width=84mm]{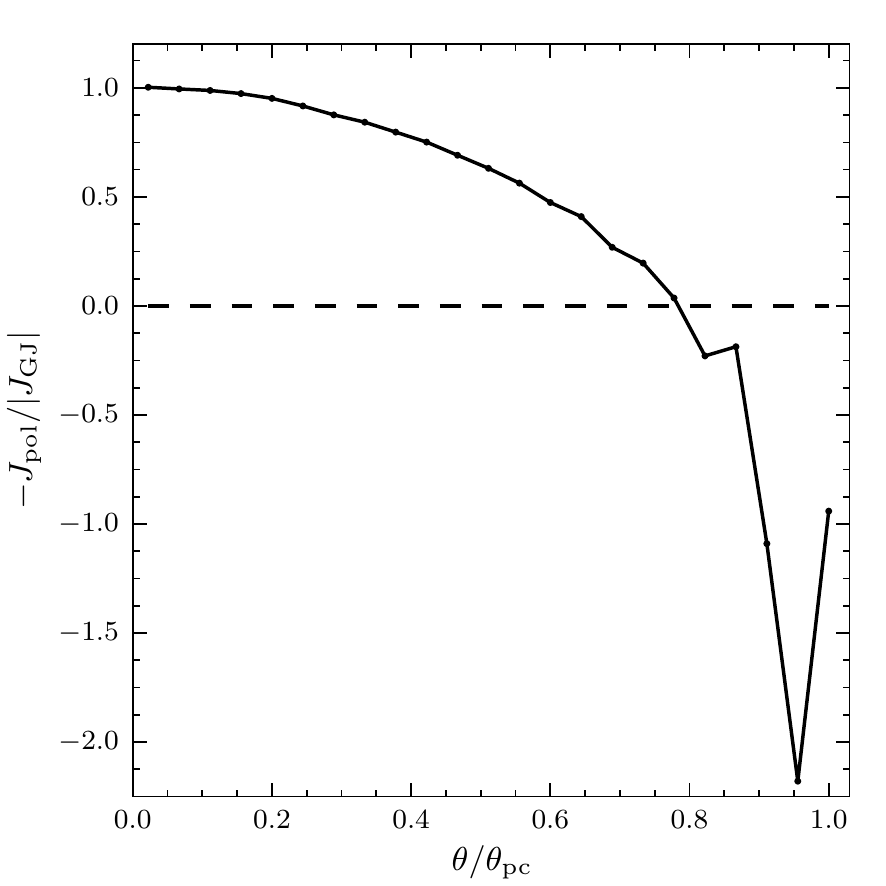}
\caption{Normalised poloidal current density across the polar cap.  $J_{\rm GJ} = \rho_{\rm GJ}$, $\rho_{\rm GJ} = - 2\, \vec{\Omega}\cdot\B$, where $c=1$ and $4\pi = 1$. $\theta_{\rm pc} = 0.088 \pi$ is the half-width of the polar cap, defined by the footpoint of the last field line in the closed zone.  \label{fig:polarcap}}
\end{figure}

\begin{figure}
\includegraphics[width=84mm]{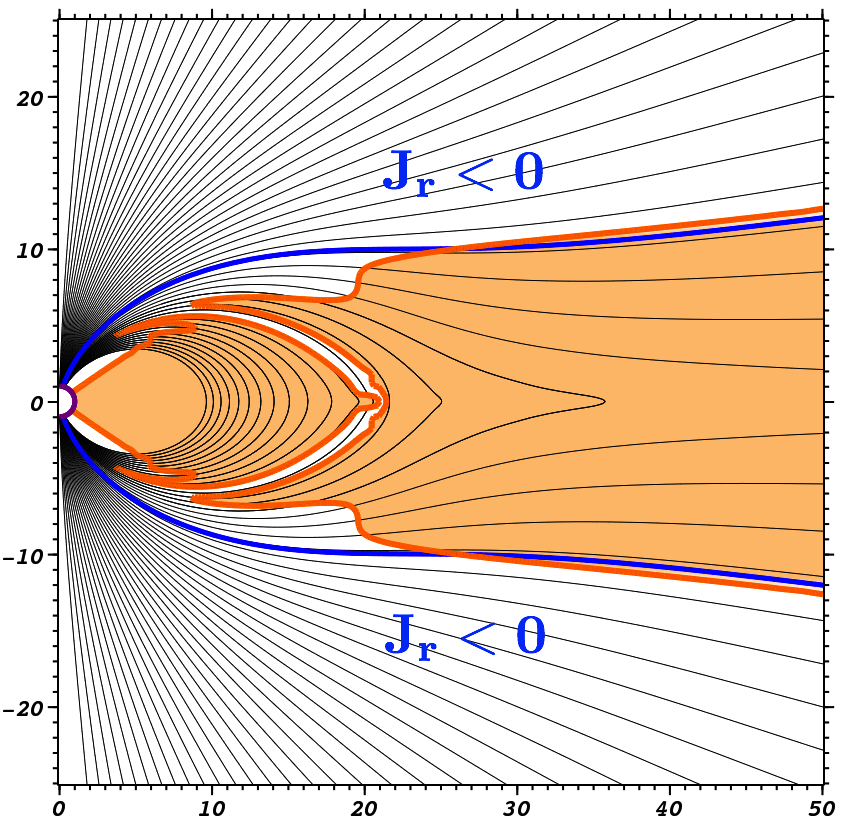}
\caption{The blue line separates the outflowing current region, with $J_r < 0$, near the poles, and the return current region nearer the equator. The orange line is the contour of zero charge density $\rho_e$, and the shaded region has $\rho_e > 0$. Poloidal field lines are drawn in black. The charge density has been cleaned with the DTV filter (Section~\ref{sec:postprocess}) to remove Gibbs oscillations near the equatorial current sheet. \label{fig:chargecontour}}
\end{figure}

\section{ Discussion }
\label{sec:discuss}

\subsection{ \textsc{PHAEDRA} }

We have found that \textsc{phaedra} converges exponentially quickly to smooth solutions, is stable and accurate when discontinuities are present, and exhibits very low numerical diffusion outside of current sheets. A critical ingredient is the full force-free current density, in particular the parallel conduction current, which maintains $\E\cdot\B=0$ without introducing any stiffness to the equations of motion. We are able to use the full current because our mesh is unstaggered, and so its evaluation doesn't require any interpolation of fields or their derivatives. The unstaggered grid is itself enabled by the inherent accuracy of the pseudospectral spatial derivatives.

The flexibility of the mapped-coordinate method allows efficient calculation in large domains, while retaining accuracy near the stellar surface at the inner boundary. It also permits resolution to be concentrated where large gradients are expected to form, a kind of fixed mesh refinement. With the adjustable parameter in the arcsine mapping of the radial Chebyshev series, a deliberate tradeoff can be made between accuracy close to the boundary and stable time step. The ability to model interactions of force-free waves with a solid boundary to high accuracy will be useful when studying, for example, turbulence driven by neutron star vibrations. The treatment of non-periodic boundaries is one of the advantages of spectral expansions over high-order finite difference methods, which can experience difficulties at surfaces.

The code is parallelised with the standard MPI library, and can be run in both shared- and distributed-memory environments. It is efficient enough to run on dense grids, using around 16--32 processors, for many millions of time steps. Simulations on coarser grids can be run on one or a few processors, on consumer laptops and workstations. One concern is that the $O(N^2)$ communication time required for the global MPI call will become a problem when scaling to a very large number of processors, for example for a three-dimensional calculation. We expect that the code should scale well up to several hundred processors on existing hardware, and that this will be sufficient.

The principal issue we have encountered is the resistivity imparted to current sheets by the spectral filters. There are numerous tricks that can be used to reduce or eliminate this in simple cases, like the aligned rotator we describe above. However, we prefer not to use any method insufficiently robust, flexible, or efficient to also be applicable to general three-dimensional current sheets propagating across the grid. The ideal aligned rotator may represent the worst-case comparison of \textsc{PHAEDRA} to codes employing finite-difference or finite-volume methods, which can evolve the solution to a steady-state with very little flux closed outside the light cylinder, either with use of a staggered mesh \citep{Spitkovsky:2006p752}, or manual nulling of the inflow of flux into the current sheet \citep{McKinney:2006p974}. We are investigating adaptive spectral- and real-space filtering, which may be able to stabilise current sheets with less dissipation than global filtering.

In any case, physical current sheets are believed to possess finite resistivity \citep[e.g.][]{2007ApJ...670..702Z, 2008ApJ...684.1477Z}. The effective hyper-resistivity of the filters confines the dissipation to grid-scale features like current sheets where it is physically expected, leaving resolved features to evolve ideally. In this sense, our aligned rotator solution may be realistic.

\subsection{ Planned Extensions }

We are planning to extend the scheme in several ways. The derivatives are already being calculated in a metric-independent fashion, and so it should be straightforward to adapt the code to the Schwarzschild or Kerr metrics, using the formalism of \citet{Komissarov:2004p747}. Eventually it should even be possible to include effects due to a time-dependent metric in this framework \citep{Komissarov:2011p4201}, possibly using data from a code which evolves the Einstein equations. 

The method currently assumes axisymmetry; we intend to make it fully three-dimensional, using either the complete `double Fourier' method on spheres or expansions in spherical harmonics. We expect the overall scheme to be adaptable to different choices of basis functions with only minor changes. This would allow us to investigate the oblique pulsar magnetosphere in spherical coordinates, and so alleviate some of the difficulties, such as stair-stepping on the inner surface and an inflexible equispaced Cartesian grid, encountered by some existing 3D force-free codes. Removing the axisymmetric restriction will also permit us to evaluate the stability of field configurations produced by general surface footpoint motions. Considering the lower resolution that is possible with 3D calculations, it is encouraging that we see close agreement between simulations on coarse and fine grids, even for the pulsar solution.

Another promising modification is to a 3D Cartesian geometry, using Fourier series in all directions. This geometry would be suitable for studying ultra-relativistic turbulent cascades \citep{Thompson:1998p4118, Cho:2005p4091} and instabilities surrounding nearly force-free current sheets. Aside from being simpler than expansions in spherical coordinates, the Cartesian-plus-Fourier combination has the benefit of allowing the solenoidal constraint on the magnetic field to be easily enforced in spectral space. 

The low intrinsic numerical diffusivity of our code makes it ideal for studying the effect of physical plasma resistivity. Although force-free electrodynamics lacks a well-defined fluid frame, and hence a preferred form for any dissipative terms, several formulations of resistive, nearly force-free, electrodynamics have been proposed \citep{Lyutikov:2003p1775, Gruzinov:2008p4525, Li:2011p4396, Kalapotharakos:2011p4587}. Being able to resolve waves with many fewer points per wavelength, spectral codes require less diffusivity than lower-order schemes to accurately transport, without unphysical oscillations, a given profile on the same grid. \citet{Brandenburg:2003p185} found that spectral derivatives permitted a viscosity five times lower than that needed for even a sixth-order finite-difference method. If enough physical resistivity is added to resolve otherwise sharp current layers, we will probably be able to dispense with the `super spectral viscosity' filtering, and the code should again achieve exponential convergence.

\section*{acknowledgments}

This work was supported in part by NASA (NNX-10-AI72G and NNX-10-AN14G), and the DOE (DE-FG02-92-ER40699). KP would like to thank Greg Bryan and Anatoly Spitkovsky for many useful discussions, and Jan Hesthaven, John Boyd, and Chi-kwan Chan for correspondence on the spectral method.

\bibliographystyle{mn2e}
\bibliography{method}

\end{document}